\documentclass[11pt,leqno]{article}

\usepackage{fullpage}
\usepackage{multirow}
\usepackage{eepic,epic}
\usepackage{epsfig}
\usepackage{graphicx}
\usepackage{color}

\usepackage{ifthen}
\usepackage{amssymb}
\usepackage{amsfonts}
\usepackage{amsmath}
\usepackage{pifont}
\usepackage{nicefrac}



\newcommand{\OMIT}[1]{} %

\newenvironment{proofs}{\noindent{\bf Proof.}\hspace*{1em}}{\literalqed\bigskip}
\def\literalqed{{\ \nolinebreak\hfill\mbox{\qedblob\quad}}}

\newcommand{\sproof}{\noindent{\bf Proof.}\hspace*{1em}}

\newcommand{\eproofof}[1]{\noindent{\hspace*{0.1in} \hfil \hfill \mbox{\literalqed{} {#1}}}\quad\bigskip}

\usepackage{ifthen}
\newcommand{\hugeDebug}{false}
\ifthenelse{\equal{\hugeDebug}{false}}{
\newcommand{\normalspacing}{\singlespacing}
}
{
\newcommand{\normalspacing}{\niceninespacing}
}

\ifthenelse{\equal{\hugeDebug}{false}}{
\pagestyle{plain}
}
{
\pagestyle{headings}
}

\makeatletter
\newcommand{\singlespacing}{\let\CS=
\@currsize\renewcommand{\baselinestretch}{1}\tiny\CS}
\newcommand{\singlespacingplus}{\let\CS=
\@currsize\renewcommand{\baselinestretch}{1.25}\tiny\CS}
\newcommand{\doublespacing}{\let\CS=
\@currsize\renewcommand{\baselinestretch}{1.75}\tiny\CS}
\newcommand{\extradoublespacing}{\let\CS=
\@currsize\renewcommand{\baselinestretch}{1.9}\tiny\CS}
\newcommand{\draftspacing}{\let\CS=
\@currsize\renewcommand{\baselinestretch}{2.0}\tiny\CS}
\newcommand{\hugedraftspacing}{\let\CS=
\@currsize\renewcommand{\baselinestretch}{2.4}\tiny\CS}
\newcommand{\niceonespacing}{\let\CS=\@currsize\renewcommand{\baselinestretch}{1.1}\tiny\CS}
\newcommand{\nicetwospacing}{\let\CS=\@currsize\renewcommand{\baselinestretch}{1.2}\tiny\CS}
\newcommand{\nicethreespacing}{\let\CS=\@currsize\renewcommand{\baselinestretch}{1.3}\tiny\CS}
\newcommand{\singlespacingplusplus}{\let\CS=\@currsize\renewcommand{\baselinestretch}{1.35}\tiny\CS}
\newcommand{\nicefourspacing}{\let\CS=\@currsize\renewcommand{\baselinestretch}{1.4}\tiny\CS}
\newcommand{\nicefivespacing}{\let\CS=\@currsize\renewcommand{\baselinestretch}{1.5}\tiny\CS}
\newcommand{\nicesixspacing}{\let\CS=\@currsize\renewcommand{\baselinestretch}{1.6}\tiny\CS}
\newcommand{\nicesevenspacing}{\let\CS=\@currsize\renewcommand{\baselinestretch}{1.7}\tiny\CS}
\newcommand{\niceeightspacing}{\let\CS=\@currsize\renewcommand{\baselinestretch}{1.8}\tiny\CS}
\newcommand{\niceninespacing}{\let\CS=\@currsize\renewcommand{\baselinestretch}{1.9}\tiny\CS}
\makeatother

\makeatletter \@beginparpenalty=10000 \makeatother

\mathcode`\0="0030      %
\mathcode`\1="0031
\mathcode`\2="0032
\mathcode`\3="0033
\mathcode`\4="0034
\mathcode`\5="0035
\mathcode`\6="0036
\mathcode`\7="0037
\mathcode`\8="0038
\mathcode`\9="0039

\newcount\hour  \newcount\minutes  \hour=\time  \divide\hour by 60
\minutes=\hour  \multiply\minutes by -60  \advance\minutes by \time
\def\mmmddyyyy{\ifcase\month\or Jan\or Feb\or Mar\or Apr\or May\or Jun\or Jul\or
  Aug\or Sep\or Oct\or Nov\or Dec\fi \space\number\day, \number\year}
\def\hhmm{\ifnum\hour<10 0\fi\number\hour :%
  \ifnum\minutes<10 0\fi\number\minutes}
\makeatletter
\def\@cite#1#2{[#1\if@tempswa , #2\fi]}
\makeatother

\makeatletter
\def\@citex[#1]#2{\if@filesw\immediate\write\@auxout{\string\citation{#2}}\fi
  \def\@citea{}\@cite{\@for\@citeb:=#2\do
    {\@citea\def\@citea{,\linebreak[0]}\@ifundefined
       {b@\@citeb}{{\bf ?}\@warning
       {Citation `\@citeb' on page \thepage \space undefined}}%
\hbox{\csname b@\@citeb\endcsname}}}{#1}}
\makeatother

\makeatletter
\def\@cite#1#2{[#1\if@tempswa , #2\fi]}
\def\@citex[#1]#2{\if@filesw\immediate\write\@auxout{\string\citation{#2}}\fi
  \def\@citea{}\@cite{\@for\@citeb:=#2\do
    {\@citea\def\@citea{,\kern1pt\linebreak[0]}\@ifundefined
       {b@\@citeb}{{\bf ?}\@warning
       {Citation `\@citeb' on page \thepage \space undefined}}%
\hbox{\csname b@\@citeb\endcsname}}}{#1}}
\makeatother

\newcommand\qedblob{\mbox{\ding{113}}}

\newcommand{\nat}{\mathbb{N}}

\newcommand{\score}{\mathit{score}}

\newtheorem{theorem}{Theorem}[section]

\newtheorem{corollary}[theorem]{Corollary}
\newtheorem{definition}[theorem]{Definition}
\newtheorem{lemma}[theorem]{Lemma}

\newcommand{\p}{\ensuremath{\mathrm{P}}}
\newcommand{\np}{\ensuremath{\mathrm{NP}}}

\title{%
The Shield that Never Was: Societies with Single-Peaked Preferences are More
Open to Manipulation and Control\thanks{%
Supported in part by DFG grants RO-1202/\{11-1, 12-1\},
NSF grants CCF-0426761, IIS-0713061, and CCF-0915792,
Polish Ministry of Science and Higher Education grant N-N206-378637,
AGH University of Science and Technology grant 11.11.120.777,
the European Science Foundation's EUROCORES program LogICCC,
and Friedrich Wilhelm Bessel Research Awards to Edith
Hemaspaandra and Lane A. Hemaspaandra.
Work done in part during visits by the first three authors to
Heinrich-Heine-Universit\"at D\"usseldorf and by the fourth author
to the University of Rochester.
Also appears as URCS-TR-2009-950.
A preliminary version of this paper appeared in the proceedings of
the 12th Conference on Theoretical Aspects of Rationality and Knowledge,
July 2009~\cite{fal-hem-hem-rot:c:single-peaked-preferences}.}}

\author{Piotr Faliszewski \\ 
        Department of Computer Science \\
        AGH University of Science and Technology \\
        30-059 Krak\'ow, Poland
\and
        Edith Hemaspaandra \\
        Department of Computer Science \\
        Rochester Institute of Technology \\
        Rochester, NY 14623, USA 
\and
        Lane A. Hemaspaandra \\ 
        Department of Computer Science \\
        University of Rochester \\
        Rochester, NY 14627, USA
\and
        J{\"o}rg Rothe \\
        Institut f\"ur Informatik \\
        Heinrich-Heine-Universit{\"a}t D{\"u}sseldorf  \\
        40225 D\"usseldorf, Germany
}

\date{September 17, 2009; revised June 19, 2010}

\begin{document}
\normalspacing\sloppy

{\singlespacing

\maketitle

}

{\centering {\bf Abstract}

}%
\begin{quotation}
{\singlespacing

\vspace*{-0.1in}
Much work has been devoted, during the past twenty years, to using
complexity to protect elections from manipulation and control.  Many
results have been obtained showing NP-hardness shields, and recently
there has been much focus on whether such worst-case hardness
protections can be bypassed by frequently correct heuristics or by
approximations.  This paper takes a very different approach: We argue
that when electorates follow the canonical political science model of
societal preferences the complexity shield never existed in the first
place.  In particular, we show that for electorates having
single-peaked preferences, many existing NP-hardness results on
manipulation and control evaporate.

} %
\end{quotation}

\section{Introduction}\label{sec:introduction}

Elections are a broadly used preference aggregation model in both human
societies and multiagent systems.
For example, elections have been proposed as a mechanism for collaborative
decision-making in such multiagent system contexts as recommender
systems/collaborative filtering~\cite{gil-hor-pen:c:collaborative-filtering}
and
planning~\cite{eph-ros:cOUTbyJOURNAL:clarke-tax,eph-ros:j:multiagent-planning}.
The importance of elections explains why
elections are studied intensely over a wide range of fields, including
political science, mathematics, social choice, artificial intelligence,
economics, and operations research.
Formally, an election system takes as input a set of candidates (or
alternatives) and a set of votes (usually each a 0-1 approval vector
over the candidates or a linear order over the candidates) and outputs a
subset of the candidates as the winner(s).

Control refers to attempts to make a given candidate
win~\cite{bar-tov-tri:j:control}
(or not win~\cite{hem-hem-rot:j:destructive-control})
an election by such participation-change actions
as adding or deleting voters or candidates.  Manipulation refers to
attempts to make a given candidate
win~\cite{bar-tov-tri:j:manipulating,bar-oli:j:polsci:strategic-voting} (or
not win~\cite{con-lan-san:j:few-candidates})
by some coalition of voters who strategically change their votes.  There is
a large literature, started by the insightful
contributions of Bartholdi, Orlin, Tovey, and
Trick~\cite{bar-oli:j:polsci:strategic-voting,bar-tov-tri:j:manipulating,bar-tov-tri:j:control}, on choosing election systems that
make control and manipulation NP-hard, i.e., on choosing
election systems that seek to make
control and manipulation computationally prohibitive
(see the survey~\cite{fal-hem-hem-rot:b:richer}).
Recently, there has been a flurry of work seeking to bypass worst-case
manipulation hardness results by frequently correct heuristics or
approximation algorithms (as a few pointers into that literature, see,
e.g.,~\cite{fri-kal-nis:c:quantiative-gib-sat,xia-con:c:generalized-scoring,xia-con:c:frequently-manipulable,con-san:c:nonexistence,pro-ros:j:juntas,bre-fal-hem-sch-sch:c:approximating-elections}).

The present paper takes a radically different approach.  We study
elections where the vote set must be ``single-peaked.'' We'll discuss
single-peakedness in more detail after stating our main contributions.
But, briefly put, single-peakedness means that there is some linear
ordering of the candidates relative to which (in the model in which votes
are linear orders) each voter's preferences always increase, always
decrease, or first increase and then decrease, or (in the model in which
votes are approval vectors) each voter's approved candidates are
contiguous within the linear order.  Single-peaked preferences,
introduced by
Black~\cite{bla:j:rationale-of-group-decision-making}
(see also \cite{bla:b:polsci:committees-elections}), model
societies that are heavily focused on one issue (taxes, war, etc.), and
the single-peaked framework is so central to political science that it has
been described as ``\emph{the} canonical setting for models of political
institutions''~\cite{gai-pat-pen:btoappear:arrow-on-single-peaked-domains}, and indeed is typically the model
of societal voting first covered in an introductory course on positive (i.e.,
theoretical) political science.
This paper's main contributions are the following:
\begin{enumerate}
\item We introduce the study of single-peakedness for approval-voting elections.
\item In Section~\ref{sec:control} we show that for both voting by approval
vectors and for voting by linear orders, many election control problems
known to be NP-hard in the general case have polynomial-time algorithms in
the single-peaked case.
\item In Section~\ref{sec:manipulation} we show that many election manipulation
problems known to be NP-hard in the general case have polynomial-time
algorithms in the single-peaked case.

However---and in this we are inspired by the path-setting work of
Walsh~\cite{wal:c:uncertainty-in-preference-elicitation-aggregation},
who showed that Single Transferable Vote (for at least 3 candidates 
and weighted votes) remains NP-hard to manipulate even in the single-peaked case---we in
Section~\ref{sec:manipulation} also show that many manipulation problems remain
NP-hard even when restricted to the single-peaked case.

\item We show, contrary to intuition and the tacit assumptions of some papers,
that even for natural systems there are cases where increasing the number of
candidates decreases the complexity. In particular, we as
Theorem~\ref{t:toodles-csl} show that in the single-peaked case
in 3-veto elections (i.e., one votes
against three candidates and for all others) manipulation is in P for up to
four candidates, is NP-hard for five candidates, and is in P for six or more
candidates.
\end{enumerate}
We mostly defer discussion of related work until after our results, as
the related work will then have more context and definitions to draw
on.  However, we mention
now a point raised by a conference referee: Since ``median voting''
and ``generalized median voter schemes'' are known to in many settings
never give a voter an incentive to vote insincerely for single-peaked
electorates (see~\cite{bar:j:strategy-proof} and
\cite{pel-sud:j:single-peakedness-and-coalition-proofness}, and
the references therein, for definitions and discussion), shouldn't
single-peaked electorates use only median voting?  Our reply is that
in real-life scenarios voting
rules are typically fixed (e.g., to be plurality or approval) and cannot
be changed when one suspects the electorate to be single-peaked. In
addition,
lack of incentive to vote insincerely regarding manipulation
does not say anything about a rule's
resistance to control. Thus, we believe that our results are very
interesting and relevant.

\section{Preliminaries}
\label{sec:prelims}

\noindent\textbf{Elections and Preferences.}\quad
An \emph{election} consists of a set $C$ of candidates and a
collection $V$ of votes.  We will consider two different models for
votes.  One is that each vote is a vector (an \emph{approval vector})
from $\{0,1\}^{\|C\|}$, denoting approval ($1$) or disapproval ($0$)
for each candidate.  The other is that each vote is a linear order (by
which we mean a strict, linear order---a complete, transitive,
antisymmetric relation) over the candidates, e.g., Bob $>$ Alice $>$
David $>$ Carol.  An \emph{election system} is a mapping that takes as
input a candidate set~$C$ and a set $V$ of votes over that candidate
set, and outputs an element of $2^C$, i.e., outputs which candidates
are \emph{winners} of the election.
(We, like 
Bartholdi, Tovey, and Trick~\cite{bar-tov-tri:j:control},
do not expressly forbid elections 
with no winners, although 
all of the many natural election systems discussed in this paper
have the property that there is always at least 
one winner when there is at least one candidate.)
Except where we explicitly state
otherwise, $V$ is a list of votes (ballots), so if three votes are the
same, they will appear three times in the list.  We will use
\emph{succinct input}~\cite{fal-hem-hem:j:bribery}
to describe the quite different input model in which each preference
that is held by one or more voters appears just once in the list and
is accompanied by a binary number stating how many voters have that
preference.

The election systems of most interest to us in this paper are the
following ones.  In \emph{approval voting}, voters vote by approval
vectors, and whichever candidate(s) get the most approvals are the
winner(s).  A \emph{scoring protocol} election, which is always
defined for a specific number $m$ of candidates, is specified by a
\emph{scoring vector}, $\alpha = (\alpha_1, \alpha_2, \ldots ,
\alpha_m) \in \nat^m$, satisfying $\alpha_1 \geq \alpha_2 \geq \dots \geq
\alpha_m$.  Votes are linear orders.  Each vote contributes
$\alpha_1$ points to that vote's most preferred candidate, $\alpha_2$
points to that vote's second most preferred candidate, and so on.  And
whichever candidate(s) get the most total points are the winner(s).
Among the most important scoring protocols for $m$ candidates are
plurality, with $\alpha = (1,\overbrace{0,\ldots,0}^{m-1})$, $j$-veto
($j \leq m$), with $\alpha =
(\overbrace{1,\ldots,1}^{m-j},\overbrace{0,\ldots,0}^{j})$, 
and Borda,
with $\alpha = (m-1, m-2, \ldots , 0)$.
For $m$ candidates,
$j$-approval and $(m-j)$-veto are the same system.  
We'll also speak of
the veto system
for an unbounded number of candidates, by which we
mean that on inputs with $m$ candidates, each voter gives $1$ point 
to
all candidates other than her least favorite candidate and $0$ 
points to her least favorite candidate.  
We'll similarly also speak of
plurality for an unbounded number of candidates, by which we
mean that on inputs with $m$ candidates, each voter gives $0$ points
to
all candidates other than her favorite candidate and $1$ 
point to her favorite candidate.  And the general case of 
approval voting is always for an unbounded number of candidates.

Let us fix some notation for the election systems defined above.  Let
$(C,V)$ be an election and let $c$ be a candidate in~$C$. By
$\score_{(C,V)}(c)$ we mean the score of $c$ in election $(C,V)$,
i.e., the number of points $c$ receives under plurality, veto, or some
given scoring protocol and the number of $c$'s approvals under
approval voting (it will always be clear from the context, which
election system the score refers to).  If the candidate set $C$ is
clear from the context, we simply write $\score_{V}(c)$.

\begin{figure}[tb!]
\centering
\setlength{\unitlength}{0.00074366in}
\begingroup\makeatletter\ifx\SetFigFontNFSS\undefined%
\gdef\SetFigFontNFSS#1#2#3#4#5{%
  \reset@font\fontsize{#1}{#2pt}%
  \fontfamily{#3}\fontseries{#4}\fontshape{#5}%
  \selectfont}%
\fi\endgroup%
{\renewcommand{\dashlinestretch}{30}
\begin{picture}(7790,3112)(0,-10)
\thicklines
\put(2287,375){\blacken\ellipse{134}{134}}
\put(2287,375){\ellipse{134}{134}}
\put(1837,375){\blacken\ellipse{134}{134}}
\put(1837,375){\ellipse{134}{134}}
\put(4132,375){\blacken\ellipse{134}{134}}
\put(4132,375){\ellipse{134}{134}}
\put(4987,375){\blacken\ellipse{134}{134}}
\put(4987,375){\ellipse{134}{134}}
\put(7237,375){\blacken\ellipse{134}{134}}
\put(7237,375){\ellipse{134}{134}}
\path(487,3075)(487,375)(7687,375)(7687,3075)
\path(487,2625)(488,2624)(491,2623)
	(497,2620)(507,2615)(520,2609)
	(538,2600)(560,2589)(588,2575)
	(621,2559)(658,2540)(700,2520)
	(747,2497)(797,2472)(851,2445)
	(908,2417)(966,2389)(1026,2359)
	(1088,2329)(1149,2299)(1210,2269)
	(1271,2239)(1332,2210)(1391,2181)
	(1448,2153)(1504,2126)(1559,2100)
	(1612,2074)(1663,2050)(1712,2026)
	(1760,2004)(1807,1982)(1852,1960)
	(1895,1940)(1938,1920)(1979,1901)
	(2019,1883)(2059,1864)(2098,1847)
	(2136,1829)(2174,1812)(2212,1796)
	(2249,1779)(2287,1762)(2326,1745)
	(2366,1728)(2405,1711)(2445,1694)
	(2485,1677)(2526,1660)(2567,1643)
	(2609,1626)(2651,1608)(2693,1591)
	(2736,1574)(2780,1556)(2823,1538)
	(2868,1521)(2913,1503)(2958,1486)
	(3003,1468)(3049,1450)(3095,1433)
	(3141,1415)(3187,1398)(3233,1381)
	(3279,1364)(3325,1347)(3371,1330)
	(3416,1314)(3461,1297)(3506,1281)
	(3551,1266)(3594,1250)(3638,1235)
	(3681,1221)(3723,1206)(3765,1192)
	(3807,1178)(3848,1164)(3889,1151)
	(3929,1138)(3969,1125)(4008,1112)
	(4048,1100)(4087,1087)(4126,1075)
	(4166,1063)(4205,1051)(4245,1039)
	(4285,1027)(4326,1015)(4367,1003)
	(4409,991)(4451,979)(4493,967)
	(4536,955)(4580,943)(4623,931)
	(4668,919)(4713,908)(4758,896)
	(4803,884)(4849,872)(4895,861)
	(4941,850)(4987,839)(5033,828)
	(5079,817)(5125,806)(5171,796)
	(5216,786)(5261,776)(5306,766)
	(5351,757)(5394,748)(5438,739)
	(5481,730)(5523,722)(5565,714)
	(5607,707)(5648,699)(5689,692)
	(5729,685)(5769,679)(5808,672)
	(5848,666)(5887,660)(5926,654)
	(5966,649)(6005,643)(6045,638)
	(6085,633)(6126,628)(6168,623)
	(6211,618)(6254,613)(6299,609)
	(6346,604)(6394,600)(6443,595)
	(6494,591)(6547,586)(6602,582)
	(6659,577)(6718,573)(6778,568)
	(6840,564)(6903,560)(6966,555)
	(7031,551)(7095,546)(7159,542)
	(7222,538)(7283,534)(7341,531)
	(7396,527)(7448,524)(7494,521)
	(7536,519)(7573,517)(7604,515)
	(7629,513)(7650,512)(7665,511)
	(7675,511)(7682,510)(7685,510)(7687,510)
\path(487,375)(489,376)(492,377)
	(499,380)(509,384)(524,390)
	(543,398)(568,408)(599,420)
	(634,435)(675,451)(721,470)
	(771,490)(825,512)(882,535)
	(942,559)(1003,583)(1066,608)
	(1129,633)(1192,659)(1254,684)
	(1316,708)(1377,732)(1436,755)
	(1493,778)(1549,800)(1603,821)
	(1655,841)(1705,861)(1754,880)
	(1801,898)(1847,915)(1892,932)
	(1935,948)(1977,964)(2018,979)
	(2059,994)(2099,1009)(2139,1023)
	(2178,1037)(2218,1051)(2257,1065)
	(2296,1079)(2336,1092)(2375,1106)
	(2415,1120)(2456,1133)(2496,1147)
	(2538,1161)(2579,1174)(2621,1188)
	(2664,1202)(2706,1216)(2750,1230)
	(2793,1244)(2837,1258)(2882,1272)
	(2926,1286)(2971,1300)(3016,1314)
	(3060,1328)(3105,1342)(3150,1356)
	(3194,1369)(3239,1383)(3283,1396)
	(3326,1409)(3369,1422)(3412,1435)
	(3453,1447)(3495,1460)(3535,1472)
	(3575,1483)(3614,1495)(3653,1506)
	(3690,1517)(3727,1528)(3763,1539)
	(3799,1549)(3834,1559)(3868,1569)
	(3901,1579)(3934,1588)(3967,1597)
	(4005,1608)(4042,1619)(4079,1630)
	(4116,1640)(4153,1650)(4190,1660)
	(4227,1670)(4264,1680)(4301,1689)
	(4339,1699)(4376,1708)(4414,1716)
	(4452,1725)(4490,1732)(4529,1740)
	(4567,1747)(4605,1753)(4643,1759)
	(4681,1765)(4719,1769)(4757,1773)
	(4794,1777)(4832,1779)(4868,1781)
	(4905,1783)(4941,1783)(4977,1783)
	(5012,1782)(5047,1780)(5082,1778)
	(5116,1775)(5151,1771)(5185,1766)
	(5219,1761)(5253,1755)(5287,1747)
	(5320,1740)(5353,1732)(5386,1723)
	(5420,1713)(5455,1703)(5490,1691)
	(5526,1679)(5562,1666)(5599,1652)
	(5637,1638)(5675,1623)(5714,1607)
	(5753,1590)(5793,1573)(5833,1555)
	(5873,1536)(5913,1517)(5954,1497)
	(5994,1477)(6035,1457)(6075,1436)
	(6115,1415)(6155,1394)(6194,1373)
	(6233,1351)(6271,1330)(6308,1309)
	(6345,1288)(6381,1267)(6417,1246)
	(6451,1225)(6485,1205)(6518,1185)
	(6551,1165)(6583,1145)(6614,1126)
	(6644,1107)(6675,1087)(6708,1066)
	(6740,1045)(6773,1024)(6805,1003)
	(6837,982)(6869,960)(6902,938)
	(6935,916)(6969,893)(7003,869)
	(7039,844)(7076,818)(7113,791)
	(7152,764)(7192,735)(7234,706)
	(7275,676)(7318,645)(7360,614)
	(7402,584)(7443,554)(7483,525)
	(7520,498)(7554,473)(7585,450)
	(7612,431)(7634,414)(7652,401)
	(7666,390)(7676,383)(7682,379)
	(7686,376)(7687,375)
\path(487,375)(488,376)(491,378)
	(496,383)(504,390)(515,400)
	(531,413)(551,430)(575,451)
	(603,476)(636,504)(674,537)
	(715,572)(760,611)(808,653)
	(858,696)(911,742)(966,789)
	(1021,837)(1078,885)(1134,933)
	(1190,982)(1246,1029)(1300,1076)
	(1354,1121)(1406,1166)(1456,1208)
	(1505,1250)(1552,1290)(1597,1328)
	(1641,1364)(1683,1399)(1723,1433)
	(1762,1465)(1799,1496)(1835,1525)
	(1869,1553)(1903,1580)(1935,1606)
	(1966,1630)(1996,1654)(2026,1677)
	(2054,1700)(2082,1721)(2110,1742)
	(2137,1762)(2173,1789)(2209,1816)
	(2244,1841)(2279,1866)(2313,1890)
	(2347,1914)(2382,1937)(2416,1960)
	(2450,1983)(2484,2005)(2518,2026)
	(2551,2047)(2585,2067)(2618,2087)
	(2651,2107)(2684,2126)(2716,2144)
	(2748,2161)(2780,2178)(2811,2195)
	(2841,2210)(2871,2225)(2900,2240)
	(2929,2254)(2957,2267)(2984,2279)
	(3011,2291)(3038,2303)(3063,2314)
	(3089,2324)(3114,2334)(3138,2344)
	(3163,2353)(3187,2362)(3215,2373)
	(3242,2383)(3270,2393)(3298,2402)
	(3327,2412)(3356,2421)(3386,2430)
	(3417,2439)(3448,2448)(3479,2456)
	(3511,2464)(3543,2472)(3576,2479)
	(3609,2485)(3642,2492)(3676,2497)
	(3709,2503)(3742,2507)(3776,2511)
	(3809,2514)(3842,2517)(3874,2519)
	(3906,2521)(3938,2521)(3970,2522)
	(4001,2521)(4032,2520)(4063,2518)
	(4094,2516)(4125,2512)(4152,2509)
	(4180,2505)(4208,2501)(4236,2496)
	(4265,2490)(4295,2484)(4326,2477)
	(4357,2470)(4388,2462)(4421,2453)
	(4454,2444)(4488,2434)(4522,2423)
	(4557,2412)(4593,2400)(4629,2387)
	(4665,2374)(4702,2360)(4739,2346)
	(4776,2332)(4813,2317)(4850,2302)
	(4887,2286)(4924,2270)(4961,2254)
	(4998,2238)(5034,2221)(5071,2205)
	(5107,2188)(5143,2171)(5179,2153)
	(5215,2136)(5251,2118)(5287,2100)
	(5318,2084)(5349,2069)(5381,2052)
	(5414,2036)(5446,2019)(5480,2001)
	(5513,1983)(5548,1965)(5583,1946)
	(5618,1927)(5654,1908)(5691,1888)
	(5728,1867)(5765,1846)(5803,1825)
	(5841,1804)(5879,1782)(5917,1760)
	(5956,1737)(5994,1715)(6033,1692)
	(6071,1669)(6109,1646)(6147,1623)
	(6184,1600)(6221,1578)(6257,1555)
	(6293,1532)(6329,1510)(6363,1487)
	(6397,1465)(6431,1444)(6463,1422)
	(6495,1400)(6527,1379)(6557,1358)
	(6587,1337)(6617,1316)(6646,1296)
	(6675,1275)(6706,1252)(6737,1229)
	(6767,1206)(6798,1183)(6828,1159)
	(6858,1134)(6889,1110)(6920,1084)
	(6952,1057)(6984,1030)(7017,1001)
	(7051,971)(7086,940)(7122,908)
	(7159,874)(7197,840)(7236,804)
	(7275,767)(7315,730)(7355,692)
	(7395,655)(7434,618)(7472,582)
	(7508,548)(7541,516)(7572,487)
	(7599,461)(7622,438)(7642,419)
	(7657,404)(7669,392)(7678,384)
	(7683,379)(7686,376)(7687,375)
\put(-200,1410){\makebox(0,0)[lb]{\smash{{\SetFigFontNFSS{10}{12.0}{\rmdefault}{\mddefault}{\updefault}Utility}}}}
\put(892,15){\makebox(0,0)[lb]{\smash{{\SetFigFontNFSS{10}{12.0}{\rmdefault}{\mddefault}{\updefault}candidate $c_1$}}}}
\put(2107,15){\makebox(0,0)[lb]{\smash{{\SetFigFontNFSS{10}{12.0}{\rmdefault}{\mddefault}{\updefault}candidate $c_2$}}}}
\put(6652,15){\makebox(0,0)[lb]{\smash{{\SetFigFontNFSS{10}{12.0}{\rmdefault}{\mddefault}{\updefault}candidate $c_5$}}}}
\put(3442,-420){\makebox(0,0)[lb]{\smash{{\SetFigFontNFSS{10}{12.0}{\rmdefault}{\mddefault}{\updefault}Position on Taxes}}}}
\put(462,-250){\makebox(0,0)[lb]{\smash{{\SetFigFontNFSS{10}{12.0}{\rmdefault}{\mddefault}{\updefault}low taxes}}}}
\put(6947,-250){\makebox(0,0)[lb]{\smash{{\SetFigFontNFSS{10}{12.0}{\rmdefault}{\mddefault}{\updefault}high taxes}}}}
\put(1387,2310){\makebox(0,0)[lb]{\smash{{\SetFigFontNFSS{10}{12.0}{\rmdefault}{\mddefault}{\updefault}$v_1$}}}}
\put(5572,2130){\makebox(0,0)[lb]{\smash{{\SetFigFontNFSS{10}{12.0}{\rmdefault}{\mddefault}{\updefault}$v_2$}}}}
\put(4537,1545){\makebox(0,0)[lb]{\smash{{\SetFigFontNFSS{10}{12.0}{\rmdefault}{\mddefault}{\updefault}$v_3$}}}}
\put(3412,15){\makebox(0,0)[lb]{\smash{{\SetFigFontNFSS{10}{12.0}{\rmdefault}{\mddefault}{\updefault}candidate $c_3$}}}}
\put(4672,15){\makebox(0,0)[lb]{\smash{{\SetFigFontNFSS{10}{12.0}{\rmdefault}{\mddefault}{\updefault}candidate $c_4$}}}}
\end{picture}
}
\vspace*{0.2in}
\caption{\label{fig:single-peaked-motivation} 
Single-Peaked Preference Motivation: Utility Curves for Three Voters.}
\end{figure}
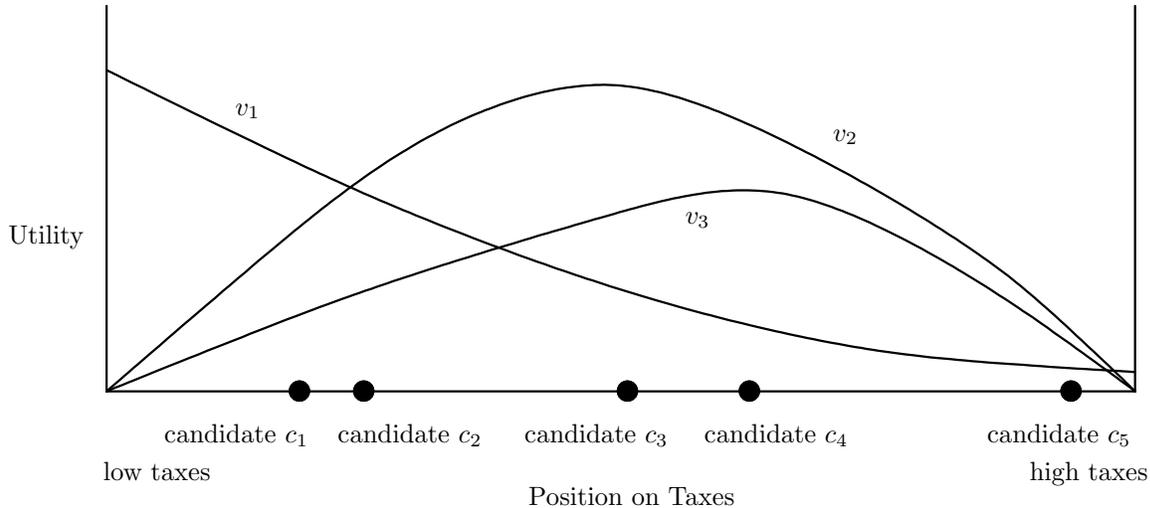

\bigskip
\noindent\textbf{Single-Peaked Preferences.}\quad
A collection $V$ of votes, each vote $v_i$ being a linear order $>_i$
over~$C$, is said to be \emph{single-peaked} exactly if there exists a
linear order over $C$, call it $L$, such that for each triple of
candidates~$c$, $d$, and~$e$, it holds that:
\[
(c\, L\, d\, L\, e \vee e\, L\, d\, L\, c) \Longrightarrow
(\forall i)\, [c\, >_i\, d \Longrightarrow d\, >_i\, e].
\]
That is just a formal way of saying that with respect to~$L$, each
voter's degree of preference rises to a peak and then falls (or just
rises or just falls).  The loose intuition behind this is captured in
Figure~\ref{fig:single-peaked-motivation}.  If we imagine an electorate
completely focused on one issue (say taxes), with each person having a
single-peaked (in the natural
analogous sense of that notion applied to curves)
utility curve (but potentially different utility curves
for different voters), one gets precisely this notion (give or take
the issue of ties).  In Figure~\ref{fig:single-peaked-motivation}'s
example, the preferences of $v_1$ would be $c_1 > c_2 > c_3 > c_4 >
c_5$, of $v_2$ would be $c_3 > c_4 > c_2 > c_1 > c_5$, and of $v_3$
would be $c_4 > c_3 > c_2 > c_1 > c_5$.

The seminal study of single-peaked preferences was done by
Black~\cite{bla:j:rationale-of-group-decision-making},
and that work and many
subsequent studies (e.g.,
\cite{nie-wri:j:voting-cycles,dav-hin-ord:j:electoral-process,poo-ros:b:congress,kre:b:pivotal-politics,bla:b:polsci:committees-elections})
argued that single-peaked preferences (in the
unidimensional spatial model) are a broadly useful model of electoral
preferences that captures many 
important settings.\footnote{\singlespacing{}We mention in passing that
since with single-peaked preferences there is always a Condorcet
winner when the number of voters is odd, every Condorcet-consistent
voting rule has an efficient winner-determination algorithm for odd
numbers of
voters~\cite{wal:c:uncertainty-in-preference-elicitation-aggregation}.
Of course, many interesting voting rules are not
Condorcet-consistent.  
}
Of course, issues
that are multidimensional---as many issues are---are typically not
captured by single-peaked preferences (by which we always mean the
unidimensional case).  And even in a society that is completely
focused on one issue, some maverick individuals may focus on other
issues, so one should keep in mind that the single-peaked case is a
widely studied but extreme model.

Looking again at Figure~\ref{fig:single-peaked-motivation}, let us
imagine that each voter has a utility threshold at which she starts
approving of candidates.  Then in the
approval-vector-as-method-of-voting we'll have that for some linear
order~$L$, each voter either disapproves of everyone or approves of
precisely a set of candidates that are contiguous with respect to~$L$.
Formally, we say a collection~$V$, made up of approval vectors $v_1,
v_2, \ldots , v_{n}$ over the candidate set~$C$, with Approves$_i$
being the set of candidates that $v_i$ approves, is
\emph{single-peaked} exactly if there is a linear order over~$C$, 
call it $L$,
such that for each triple of
candidates~$c$, $d$, and~$e$, it holds that:
\[
c\, L\, d\, L\, e  \Longrightarrow
(\forall i)\, [\{c,e\} \subseteq \textrm{Approves}_i \Longrightarrow 
d \in \textrm{Approves}_i].
\]
The reasons that single-peaked approval voting is compellingly
natural to study (and the reasons why one should not go overboard
and claim that it is a universally appropriate notion) are essentially
the same as those, touched on above, 
involving single-peaked voting with respect 
to preference orders.  In many cases it is natural to assume that there
is some unidimensional issue steering society and that each 
person's range of comfort on that issue is some contiguous 
segment along that issue's dimension.  In such a case, each 
person's set of approved candidates
will form a contiguous block among the candidates when they are 
ordered by their positions 
on that issue.
To the best of our knowledge, this is the first paper to study
elections based on
single-peaked approval vectors---although the literature is so vast
that we would not be at all surprised 
such elections had been previously studied.

In our control and manipulation problems, we'll for the single-peaked
case follow
Walsh~\cite{wal:c:uncertainty-in-preference-elicitation-aggregation}
and take as part of the input a particular linear order of the candidates
relative to which the votes are single-peaked.  This is arguably
natural---as we may view candidates' positions on the issue that
defines the entire election as being openly known.  However, one may
wonder how hard it is to, given a set of voters, tell whether it is
single-peaked.  For the case of votes being linear orders, Bartholdi
and Trick (\cite{bar-tri:j:stable-matching-from-psychological-model},
see also~\cite{doi-fal:j:unidimensional,esc-lan-ozt:c:single-peaked-consistency};
a somewhat related paper 
is~\cite{tri:j:single-peaked-preferences-on-a-tree}) show by a
path-based graph algorithm that the problem is in $\p$, and 
Doignon and Falmagne~\cite{doi-fal:j:unidimensional} and 
Escoffier, Lang, and
{\"O}zt{\"u}rk~\cite{esc-lan-ozt:c:single-peaked-consistency}
show
how to produce in polynomial time a linear order witnessing the
single-peakedness when such an ordering
exists.
For the case of
votes being approval vectors, 
the literature already has long 
contained the analogue of both the results just mentioned.
In particular, the work of 
Fulkerson and Gross~\cite[Sections~5 and~6]{ful-gro:j:contiguous-ones}
and 
Booth and Lueker~\cite[Theorem~6]{boo-lue:j:consecutive-ones-property}
proves (rephrased in our terminology) that in 
polynomial time---in a certain natural sense, even in linear time---one
can determine whether a set of approval vectors is single-peaked,
can in such a case efficiently find a linear ordering realizing the 
single-peakedness, and 
indeed can implicitly represent all linear orderings
realizing the single-peakedness.

\begin{theorem}[\cite{ful-gro:j:contiguous-ones,boo-lue:j:consecutive-ones-property}]
\label{t:embed}
Given a collection $V$ of approval vectors over~$C$, in polynomial
time we can produce a linear order $L$ witnessing $V$'s
single-peakedness or can determine that $V$ is not single-peaked.
\end{theorem}

Note that Theorem~\ref{t:embed} does not seem to follow from
the analogous result for linear orders.
For example, extending each approval vector to
form a linear order with all approved candidates preceding all the
disapproved ones, and then running the algorithm
of Escoffier, Lang, and
{\"O}zt{\"u}rk~\cite{esc-lan-ozt:c:single-peaked-consistency} on the
thus transformed input does not work.
\bigskip
\noindent\textbf{Control and Manipulation Problems.}\quad
For an election system
$\mathcal{E}$, the \emph{Constructive Control by Adding Candidates}
problem is the set of all $(C,V,p,k,C')$, where $V$
consists of votes over $C
\cup C'$, $p \in C$, and $C \cap C' = \emptyset$, such that there is a
set $C'' \subseteq C'$ with $\|C''\| \leq k$ for which candidate $p$
is the unique winner under election system $\mathcal{E}$ when the
voters in $V$ vote over $C \cup C''$ (i.e., we restrict each voter
down to her induced preferences over $C \cup C''$).  The
\emph{Destructive Control by Adding Candidates} problem is the same
except now the question is whether there is such a $C''$ ensuring that
$p$ is \emph{not} a unique winner.

The Constructive/Destructive Control by Deleting Candidates problems
are analogous, with inputs of the form $(C,V,p,k)$, where $k$ is a
limit on how many candidates from $C$ can be deleted, and it is
forbidden to delete~$p$.  The Constructive/Destructive
Control by Adding/Deleting
Voters problems are analogous, with inputs respectively $(C,V,p,k,V')$
and $(C,V,p,k)$, where $V'$, $V' \cap V = \emptyset$, is a pool of
``unregistered'' voters, and $k$ denotes the bound on how many voters
from $V'$ we can add (Adding case) or how many voters from $V$ we can
delete (Deleting case).

\emph{Constructive/Destructive Control by Unlimited Adding Candidates}
is the same as Constructive/Destructive Control by Adding Candidates
except there is no ``$k$''; one may add any or all of the members
of~$C'$.

Most of these problems were introduced in the seminal control paper of
Bartholdi, Tovey, and Trick~\cite{bar-tov-tri:j:control}, and the
remaining ones were introduced
in~\cite{hem-hem-rot:j:destructive-control,fal-hem-hem-rot:j:llull}.
These problems model such real-world problems as introducing a
candidate to run to split off another candidate's support; urging 
an independent candidate to withdraw for the good of the country;
spreading rumors that people with outstanding warrants who try to vote
will be arrested; and sending vans to retirement homes to drive
car-less members of one's party to the voting place.

In each case, the inputs we specified are for the ``general'' case.
For the single-peaked case there will be an additional input, a linear
order $L$ over the candidates such that relative to $L$ the election
is single-peaked (with respect to all voters and candidates---even
those in $C'$ and in $V'$ in problems having those extra sets).  (If
$L$ is not such a linear order, the input is immediately rejected.
$L$'s
goodness can be easily tested in polynomial time, simply
by looking at each vote.)  As mentioned previously,
in assuming that $L$ is given we are following
Walsh's~\cite{wal:c:uncertainty-in-preference-elicitation-aggregation}
natural model---$L$ is the broadly known positioning of the
candidates.\footnote{\singlespacing{}Both for linear-order votes
  (\cite{doi-fal:j:unidimensional}, see
  also~\cite{bar-tri:j:stable-matching-from-psychological-model,esc-lan-ozt:c:single-peaked-consistency}) 
  and
  for approval-vector votes (Theorem~\ref{t:embed}) it holds that
  even if $L$ is not given one can find a good $L$ in polynomial time
  if one exists.  This fact may be comforting to those who would
  prefer that $L$ not be given.  But we caution that a given vote set 
  $V$ may have many valid $L$'s.  And for some problems, which valid $L$
  one uses may affect the problem and its complexity.  Indeed, not
  having $L$ as part of the input might even open the door to
  time-sequence issues, e.g., in deletion of candidates/voters,
  should we ask instead whether single-peakedness should only have to
  hold after the deletion?  (We'd say, ideally, ``no.'')  These issues
  are reasonable for further control/manipulation-related study (and
  the issue of whether $L$ is known \emph{a priori} is much studied in
  the political economy literature already, see, e.g.,
  \cite[Section~2.4]{aus-ban:b:positive-political-II}), but we find
  the
  Walsh~\cite{wal:c:uncertainty-in-preference-elicitation-aggregation}
  model to be most natural and compelling.  In many cases our proofs
  work fine if $L$ is not part of the input, and at times we'll
  mention that.}

The (constructive coalition weighted) manipulation problem was
introduced by Conitzer, Sandholm, and Lang
(\cite{con-lan-san:j:few-candidates} and its conference precursors,
building on the seminal manipulation
papers~\cite{bar-oli:j:polsci:strategic-voting,bar-tov-tri:j:manipulating})
and takes as input a list of candidates~$C$, a list of nonmanipulative
voters each specified by preferences (as a linear order or as an
approval vector, depending on our election system) over $C$ and a
nonnegative
integer weight, a list of the weights of the voters in our
manipulating coalition, and the candidate $p$ that our coalition seeks
to make a winner.  The set of all such inputs for which there is some
assignment of preferences to the manipulators that makes $p$ a winner
is the \emph{Constructive Coalition Weighted Manipulation} problem for
that election system.  The \emph{Constructive Size-$k$-Coalition
  Unweighted Manipulation} problem is defined analogously, except with
all voters having unit weight and with $k$ being the number of
manipulators.

We follow the convention
from~\cite{bar-tov-tri:j:manipulating,bar-oli:j:polsci:strategic-voting,bar-tov-tri:j:control}
of focusing on the unique-winner case for control and the winner case
(i.e., asking whether the candidate can be made \emph{a}
winner or be prevented from 
being \emph{a} winner;  following 
the literature, we'll sometimes refer to that
as the ``nonunique-winner model/case'')
for manipulation.  In many cases we've shown a given result in both
models and will sometimes mention that in passing.  All of the results
of Conitzer, Sandholm, and Lang of relevance to us hold in both models
(see Footnote~7 of~\cite{con-lan-san:j:few-candidates}).  

For the
single-peaked case, a linear order $L$ is given as part of the input
and the manipulating coalition's votes must all be single-peaked with
respect to $L$ (as must all other voters) for the input to be
accepted.

\bigskip
\noindent\textbf{Complexity Notions for Control and Manipulation Problems.}\quad
If by a given type of constructive control
action we can never change $p$ from not unique winning (not winning)
to unique winning (winning) we say the problem is \emph{immune} to
that control type, and otherwise we call it
\emph{susceptible} to that control type.
The destructive case is analogous.  If a
susceptible problem is in~$\p$ we call it \emph{vulnerable}, and if it
is $\np$-hard we call it \emph{resistant}.
A manipulation problem (such as the constructive coalition weighted
manipulation problem defined above) is said to be
\emph{vulnerable} if it is in P and \emph{resistant} if it is NP-hard.
In each control/manipulation case in this paper
where we assert vulnerability (or 
membership in P), 
not only is the decision problem in
$\p$ but also in polynomial time we can produce a
successful control/manipulation action if one 
exists (i.e., the problem is
what~\cite{hem-hem-rot:j:destructive-control} calls \emph{certifiably
  vulnerable}).  
Most of these notions are
in detail from or in the general spirit of the work of
Bartholdi,
Orlin, Tovey, and Trick
\cite{bar-tov-tri:j:control,bar-tov-tri:j:manipulating,bar-oli:j:polsci:strategic-voting},
as in some cases naturally modified or extended 
in~\cite{hem-hem-rot:j:hybrid,hem-hem-rot:j:destructive-control}.

Almost all $\np$-hardness proofs of this paper follow via reductions
from the well-known $\np$-complete problem PARTITION (see,
e.g.,~\cite{gar-joh:b:int}). Specifically, we use the variant of the
problem where our input contains a set $\{k_1,\ldots, k_n\}$ of $n$ distinct
positive integers that sum to $2K$, and we ask: Does there exist a
subset $A \subseteq \{1, \ldots, n\}$ such that $\sum_{i \in A}k_i =
K$?

\section{Control}\label{sec:control}

\subsection{Results}

We now turn to our results on control of elections. The theme of this
paper is that electorates limited to being single-peaked often are simpler
to control and manipulate.  Intuitively speaking, we will show that the
more limited range of vote collections allowed by single-peaked voting (as
opposed to general voting) is so restrictive that the reductions
showing NP-hardness fall apart---that those reductions centrally need complex
collections of votes to work.  Of course, for all we know, perhaps P = NP,
and so attempting to show that no reduction at all can exist from SAT
to our single-peaked control problems would be a fool's errand (or a Turing
Award-scale quest).  Rather, we'll show our single-peaked control problems
easy the old-fashioned way: We'll prove that they are in P.
The techniques we use to prove them in P vary from easy observations to
smart greedy schemes to dynamic programming.

The control complexity of approval voting is studied in detail by
Hemaspaandra, Hemaspaandra, and
Rothe~\cite{hem-hem-rot:j:destructive-control} (see~\cite{erd-now-rot:j:sp-av}
regarding limited adding of
candidates, see also the
survey~\cite{bau-erd-hem-hem-rot:btoappear-with-TR-pointer:computational-apects-of-approval-voting}
for an overview of computational properties of approval voting).  Among all
the adding/deleting control cases 
defined in this paper (ten in total), precisely two  are resistant, and the
other eight (all five destructive cases and the three constructive
candidate cases) are immune or vulnerable.  The two resistant cases are
Constructive Control by Adding Voters and
Constructive Control by Deleting Voters.

The following theorem shows that both of these complexity shields
evaporate for societies with single-peaked preferences.
Briefly put,
the challenge here is that the set $V'$
of voters to add may be filled with ``incomparable'' voter pairs---pairs
such that regarding their interval with respect to the linear order
defining single-peakedness, neither is a subset of the other---and so  it
is not immediately obvious what voters to add. We solve this by a ``smart
greedy'' approach, breaking votes first into broad groups based on
where their intervals' right edges fall with respect to just a certain
``dangerous'' subset of the candidates, and then re-sort those based on
their left edges, and we argue that if any strategy will reach the control
goal then this one will.

\begin{theorem}\label{t:approval-control}
For the single-peaked case, approval voting is vulnerable to constructive
control by adding voters and to  constructive
control by deleting voters, in both the unique-winner model and
the nonunique-winner model, for both the standard input model and the
succinct input model.\footnote{\singlespacing{}This result holds in our settled model in which the linear order specifying
the society's order on the candidates is part of the input, and also holds
in the model
in which the linear order is not part of the input but rather 
the question is whether there exists any valid 
linear order relative to which there is a way of 
achieving our control goal.}
\end{theorem}

\begin{proofs}
We'll cover the unique-winner cases only.  From these
it will 
be completely clear how to handle the nonunique-winner cases.  
We'll cover just the succinct-input cases, as each implies 
the corresponding standard-input case.  Susceptibility is immediately 
clear by easy examples.

First, let's speak of the societal linear order. In the model 
in which the (purported) linear order is part of the input, start by doing 
the obvious polynomial-time check to see whether this linear order 
truly is valid with respect to the voters $V$, the unregistered
voters $V'$, and the candidates $C$.  That is, check to ensure 
that each member of $V$ and $V'$ approves either of no candidates
or approves of a exactly a collection of candidates that are 
contiguous with respect to the input linear order.
If so, the votes of $V \cup V'$, which are approval vectors 
over $C$, are single-peaked with respect to the input societal 
linear order over $C$.  If not, output ``Invalid societal linear
order'' and halt.
(This very easy observation in effect shows that, for 
the approval-vector case, 
checking a given purported societal linear order is in 
P\@.  That is equally obviously true for the votes-are-linear-orders
setting, of course.)

If we're in the model in which a linear order must be found, if one
exists, relative to which the problem is single-peaked and the given
control action is possible, simply use 
Theorem~\ref{t:embed} to efficiently 
find some linear order relative to which our
input is single-peaked.  If none exists, state that and terminate.  If
Theorem~\ref{t:embed} gives us some valid order, it is not
hard to see that the algorithm we are going to give in this proof will
find a successful control action if and only if there exists any valid
societal linear 
order relative to
which a successful control action exists.  That is basically
because if $\pi_1$ and $\pi_2$ are valid 
societal linear orders, the votes are
the same either way, and in each case the algorithm we are about to
give will succeed exactly if there exists a legal-cardinality
collection of unregistered voters that achieves the control goal.

\begin{figure}[tb!]
\centering
\setlength{\unitlength}{0.00054366in}
\begingroup\makeatletter\ifx\SetFigFont\undefined%
\gdef\SetFigFont#1#2#3#4#5{%
  \reset@font\fontsize{#1}{#2pt}%
  \fontfamily{#3}\fontseries{#4}\fontshape{#5}%
  \selectfont}%
\fi\endgroup%
{\renewcommand{\dashlinestretch}{30}
\begin{picture}(10162,7035)(0,-10)
\thicklines
\put(2940,2085){\blacken\ellipse{90}{90}}
\put(2940,2085){\ellipse{90}{90}}
\put(4740,2085){\blacken\ellipse{90}{90}}
\put(4740,2085){\ellipse{90}{90}}
\put(3390,1860){\blacken\ellipse{90}{90}}
\put(3390,1860){\ellipse{90}{90}}
\put(5190,1860){\blacken\ellipse{90}{90}}
\put(5190,1860){\ellipse{90}{90}}
\put(2040,1635){\blacken\ellipse{90}{90}}
\put(2040,1635){\ellipse{90}{90}}
\put(6990,1635){\blacken\ellipse{90}{90}}
\put(6990,1635){\ellipse{90}{90}}
\put(4740,1410){\blacken\ellipse{90}{90}}
\put(4740,1410){\ellipse{90}{90}}
\put(6090,1410){\blacken\ellipse{90}{90}}
\put(6090,1410){\ellipse{90}{90}}
\put(2940,1185){\blacken\ellipse{90}{90}}
\put(2940,1185){\ellipse{90}{90}}
\put(6090,1185){\blacken\ellipse{90}{90}}
\put(6090,1185){\ellipse{90}{90}}
\put(4290,960){\blacken\ellipse{90}{90}}
\put(4290,960){\ellipse{90}{90}}
\put(5640,960){\blacken\ellipse{90}{90}}
\put(5640,960){\ellipse{90}{90}}
\put(2490,735){\blacken\ellipse{90}{90}}
\put(2490,735){\ellipse{90}{90}}
\put(2940,735){\blacken\ellipse{90}{90}}
\put(2940,735){\ellipse{90}{90}}
\put(5190,510){\blacken\ellipse{90}{90}}
\put(5190,510){\ellipse{90}{90}}
\put(5640,510){\blacken\ellipse{90}{90}}
\put(5640,510){\ellipse{90}{90}}
\path(1590,6585)(1590,2535)(10140,2535)
\path(2355,2535)(2355,4335)(2625,4335)(2625,2535)
\path(2805,2535)(2805,2985)(3075,2985)(3075,2535)
\path(4155,2535)(4155,2985)(4425,2985)(4425,2535)
\path(4605,2535)(4605,3435)(4875,3435)(4875,2535)
\path(5055,2535)(5055,2985)(5325,2985)(5325,2535)
\path(5505,2535)(5505,2985)(5775,2985)(5775,2535)
\path(6405,2535)(6405,2985)(6675,2985)(6675,2535)
\path(6855,2535)(6855,3885)(7125,3885)(7125,2535)
\path(7755,2535)(7755,3885)(8025,3885)(8025,2535)
\path(8205,2535)(8205,4335)(8475,4335)(8475,2535)
\path(8655,2535)(8655,3435)(8925,3435)(8925,2535)
\path(9555,2535)(9555,4785)(9825,4785)(9825,2535)
\path(2940,2085)(4740,2085)
\path(3390,1860)(5190,1860)
\path(2040,1635)(6990,1635)
\path(4740,1410)(6090,1410)
\path(2940,1185)(6090,1185)
\path(4290,960)(5640,960)
\path(2490,735)(2940,735)
\path(5190,510)(5640,510)
\thinlines
\dashline{60.000}(2040,2220)(2040,1635)
\dashline{60.000}(2490,2220)(2490,735)
\dashline{60.000}(2940,2220)(2940,735)
\dashline{60.000}(3390,2220)(3390,1860)
\dashline{60.000}(4290,2220)(4290,960)
\dashline{60.000}(4740,2220)(4740,1410)
\dashline{60.000}(5190,2220)(5190,510)
\dashline{60.000}(5640,2220)(5640,510)
\dashline{60.000}(6090,2220)(6090,1185)
\dashline{60.000}(6990,2220)(6990,1635)
\thicklines
\texture{55888888 88555555 5522a222 a2555555 55888888 88555555 552a2a2a 2a555555 
	55888888 88555555 55a222a2 22555555 55888888 88555555 552a2a2a 2a555555 
	55888888 88555555 5522a222 a2555555 55888888 88555555 552a2a2a 2a555555 
	55888888 88555555 55a222a2 22555555 55888888 88555555 552a2a2a 2a555555 }
\shade\path(1905,2535)(1905,5685)(2175,5685)
	(2175,2535)(1905,2535)
\path(1905,2535)(1905,5685)(2175,5685)
	(2175,2535)(1905,2535)
\shade\path(9105,2535)(9105,5685)(9375,5685)
	(9375,2535)(9105,2535)
\path(9105,2535)(9105,5685)(9375,5685)
	(9375,2535)(9105,2535)
\shade\path(3255,2535)(3255,4335)(3525,4335)
	(3525,2535)(3255,2535)
\path(3255,2535)(3255,4335)(3525,4335)
	(3525,2535)(3255,2535)
\shade\path(7305,2535)(7305,4335)(7575,4335)
	(7575,2535)(7305,2535)
\path(7305,2535)(7305,4335)(7575,4335)
	(7575,2535)(7305,2535)
\shade\path(3705,2535)(3705,3435)(3975,3435)
	(3975,2535)(3705,2535)
\path(3705,2535)(3705,3435)(3975,3435)
	(3975,2535)(3705,2535)
\shade\path(5955,2535)(5955,3885)(6225,3885)
	(6225,2535)(5955,2535)
\path(5955,2535)(5955,3885)(6225,3885)
	(6225,2535)(5955,2535)
\thinlines
\dottedline{45}(1590,2985)(10140,2985)
\dottedline{45}(1590,3435)(10140,3435)
\dottedline{45}(1590,3885)(10140,3885)
\dottedline{45}(1590,4335)(10140,4335)
\dottedline{45}(1590,4785)(10140,4785)
\dottedline{45}(1590,5235)(10140,5235)
\dottedline{45}(1590,5685)(10140,5685)
\dottedline{45}(1590,6135)(10140,6135)
\put(1365,2947){\makebox(0,0)[lb]{\smash{{\SetFigFont{10}{12.0}{\rmdefault}{\mddefault}{\updefault}$1$}}}}
\put(1365,3397){\makebox(0,0)[lb]{\smash{{\SetFigFont{10}{12.0}{\rmdefault}{\mddefault}{\updefault}$2$}}}}
\put(1365,3847){\makebox(0,0)[lb]{\smash{{\SetFigFont{10}{12.0}{\rmdefault}{\mddefault}{\updefault}$3$}}}}
\put(1365,4297){\makebox(0,0)[lb]{\smash{{\SetFigFont{10}{12.0}{\rmdefault}{\mddefault}{\updefault}$4$}}}}
\put(1365,4792){\makebox(0,0)[lb]{\smash{{\SetFigFont{10}{12.0}{\rmdefault}{\mddefault}{\updefault}$5$}}}}
\put(1365,5197){\makebox(0,0)[lb]{\smash{{\SetFigFont{10}{12.0}{\rmdefault}{\mddefault}{\updefault}$6$}}}}
\put(1365,5647){\makebox(0,0)[lb]{\smash{{\SetFigFont{10}{12.0}{\rmdefault}{\mddefault}{\updefault}$7$}}}}
\put(1365,6097){\makebox(0,0)[lb]{\smash{{\SetFigFont{10}{12.0}{\rmdefault}{\mddefault}{\updefault}$8$}}}}
\put(100,4300){\makebox(0,0)[lb]{\smash{{\SetFigFont{10}{12.0}{\rmdefault}{\mddefault}{\updefault}from $V$}}}}
\put(100,4630){\makebox(0,0)[lb]{\smash{{\SetFigFont{10}{12.0}{\rmdefault}{\mddefault}{\updefault}approvals}}}}
\put(100,4900){\makebox(0,0)[lb]{\smash{{\SetFigFont{10}{12.0}{\rmdefault}{\mddefault}{\updefault}number of}}}}
\put(1975,2317){\makebox(0,0)[lb]{\smash{{\SetFigFont{10}{12.0}{\rmdefault}{\mddefault}{\updefault}$\ell_6$}}}}
\put(2425,2317){\makebox(0,0)[lb]{\smash{{\SetFigFont{10}{12.0}{\rmdefault}{\mddefault}{\updefault}$\ell_5$}}}}
\put(2875,2317){\makebox(0,0)[lb]{\smash{{\SetFigFont{10}{12.0}{\rmdefault}{\mddefault}{\updefault}$\ell_4$}}}}
\put(3325,2317){\makebox(0,0)[lb]{\smash{{\SetFigFont{10}{12.0}{\rmdefault}{\mddefault}{\updefault}$\ell_3$}}}}
\put(3775,2317){\makebox(0,0)[lb]{\smash{{\SetFigFont{10}{12.0}{\rmdefault}{\mddefault}{\updefault}$\ell_2$}}}}
\put(4225,2317){\makebox(0,0)[lb]{\smash{{\SetFigFont{10}{12.0}{\rmdefault}{\mddefault}{\updefault}$\ell_1$}}}}
\put(4675,2317){\makebox(0,0)[lb]{\smash{{\SetFigFont{10}{12.0}{\rmdefault}{\mddefault}{\updefault}$p$}}}}
\put(5125,2317){\makebox(0,0)[lb]{\smash{{\SetFigFont{10}{12.0}{\rmdefault}{\mddefault}{\updefault}$r_1$}}}}
\put(5575,2317){\makebox(0,0)[lb]{\smash{{\SetFigFont{10}{12.0}{\rmdefault}{\mddefault}{\updefault}$r_2$}}}}
\put(6025,2317){\makebox(0,0)[lb]{\smash{{\SetFigFont{10}{12.0}{\rmdefault}{\mddefault}{\updefault}$r_3$}}}}
\put(6475,2317){\makebox(0,0)[lb]{\smash{{\SetFigFont{10}{12.0}{\rmdefault}{\mddefault}{\updefault}$r_4$}}}}
\put(6925,2317){\makebox(0,0)[lb]{\smash{{\SetFigFont{10}{12.0}{\rmdefault}{\mddefault}{\updefault}$r_5$}}}}
\put(7375,2317){\makebox(0,0)[lb]{\smash{{\SetFigFont{10}{12.0}{\rmdefault}{\mddefault}{\updefault}$r_6$}}}}
\put(7825,2317){\makebox(0,0)[lb]{\smash{{\SetFigFont{10}{12.0}{\rmdefault}{\mddefault}{\updefault}$r_7$}}}}
\put(8275,2317){\makebox(0,0)[lb]{\smash{{\SetFigFont{10}{12.0}{\rmdefault}{\mddefault}{\updefault}$r_8$}}}}
\put(8725,2317){\makebox(0,0)[lb]{\smash{{\SetFigFont{10}{12.0}{\rmdefault}{\mddefault}{\updefault}$r_9$}}}}
\put(9145,2317){\makebox(0,0)[lb]{\smash{{\SetFigFont{10}{12.0}{\rmdefault}{\mddefault}{\updefault}$r_{10}$}}}}
\put(9595,2317){\makebox(0,0)[lb]{\smash{{\SetFigFont{10}{12.0}{\rmdefault}{\mddefault}{\updefault}$r_{11}$}}}}
\put(-215,1027){\makebox(0,0)[lb]{\smash{{\SetFigFont{10}{12.0}{\rmdefault}{\mddefault}{\updefault}votes in $V' \quad \left\{\begin{array}{l} \\ \\ \\ \\ \\ \\ \end{array}\right.$}}}}
\put(4020,22){\makebox(0,0)[lb]{\smash{{\SetFigFont{10}{12.0}{\rmdefault}{\mddefault}{\updefault}etc.}}}}
\put(4965,2007){\makebox(0,0)[lb]{\smash{{\SetFigFont{10}{12.0}{\rmdefault}{\mddefault}{\updefault}$1$}}}}
\put(5415,1782){\makebox(0,0)[lb]{\smash{{\SetFigFont{10}{12.0}{\rmdefault}{\mddefault}{\updefault}$2$}}}}
\put(7215,1557){\makebox(0,0)[lb]{\smash{{\SetFigFont{10}{12.0}{\rmdefault}{\mddefault}{\updefault}$4$}}}}
\put(6315,1332){\makebox(0,0)[lb]{\smash{{\SetFigFont{10}{12.0}{\rmdefault}{\mddefault}{\updefault}$7$}}}}
\put(6315,1107){\makebox(0,0)[lb]{\smash{{\SetFigFont{10}{12.0}{\rmdefault}{\mddefault}{\updefault}$3$}}}}
\put(5865,882){\makebox(0,0)[lb]{\smash{{\SetFigFont{10}{12.0}{\rmdefault}{\mddefault}{\updefault}$1$}}}}
\put(3165,657){\makebox(0,0)[lb]{\smash{{\SetFigFont{10}{12.0}{\rmdefault}{\mddefault}{\updefault}$9$}}}}
\put(5865,432){\makebox(0,0)[lb]{\smash{{\SetFigFont{10}{12.0}{\rmdefault}{\mddefault}{\updefault}$5$}}}}
\end{picture}
}
\caption{\label{f:lane2} Bar chart of approvals from $V$, plus
intervals indicating the votes in $V'$ and their multiplicities.  The
candidates are ordered in the societal linear order and the
``dangerous'' candidates are marked with shaded bars.}
\end{figure}
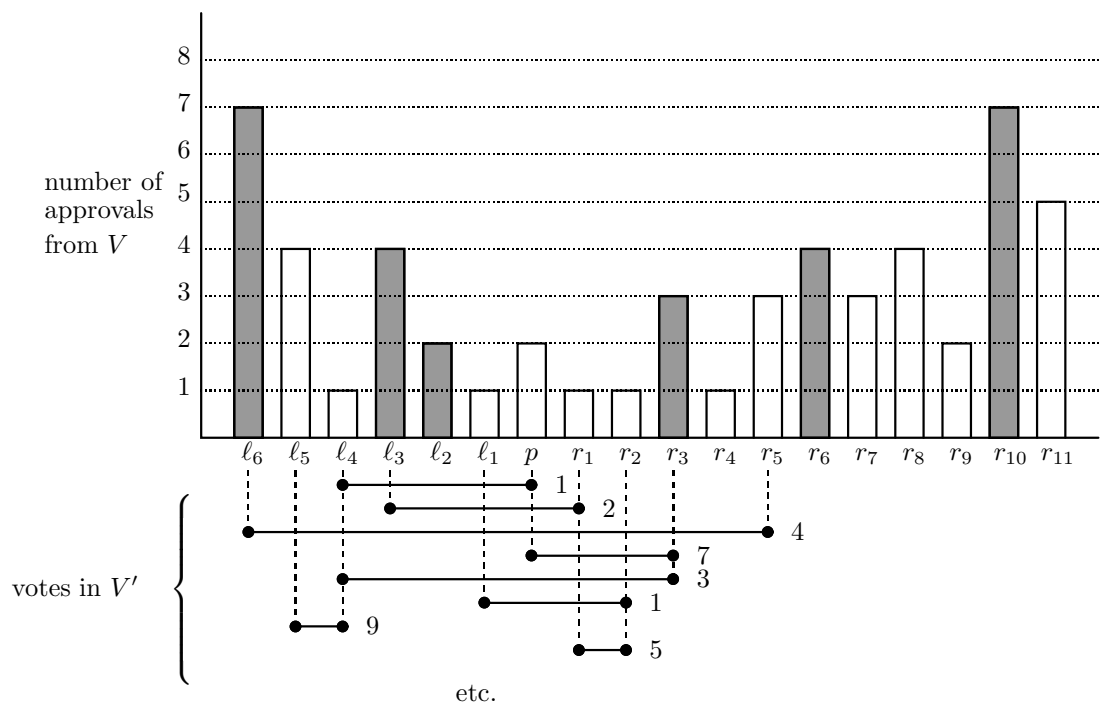

So, let us take it that we have the societal linear order, which we'll
refer to as $L$.  We start with the case of adding voters.
This proof will be done rather visually, and
although we'll implicitly be sketching a general proof, we'll keep
close to heart an example, as that will make the idea of the proof
clearer.  Let $p$ be the preferred candidate---the one we seek to make
a unique winner.  Figure~\ref{f:lane2} shows a representation of the
initial state.  In this figure, we have (for display purposes) renamed
all the candidates in a way that echos the societal linear order $L$,
i.e., the candidates are each labeled with respect to whether they are
left or right of $p$ in $L$, and 
are ordered following the societal 
order.  The bars represent the total number of
approvals each candidate gets over all the votes contained in $V$.
(Since we're in the succinct-input case, we of course do in computing
that take into account the multiplicities, e.g., if candidate $r_{10}$
is approved of in one ordering that occurs with multiplicity 4 and one
that occurs with multiplicity 3, and in no others, then the bar for 
$r_{10}$
would show that it gets 7 (not 2) approvals.)  The intervals below the
bar graph reflect the votes in $V'$.  
Note that each vote type in $V'$
either approves of no candidates (in which case we won't even display it;
it is of no help toward meeting our control goal) or, necessarily,
is an interval (although perhaps a degenerate one that contains exactly
one candidate) with respect to $L$.  
(Since we're handling the 
succinct case, votes of the same vote type are in the 
figure listed together 
along with their multiplicity to their right.
We do this because
in the input to the problem votes come in by type 
and with a listed-in-binary multiplicity, in 
the succinct case.)

The number of vote types is $O(\| C \|^2)$, since each
vote type (except for disapproving of everyone) is defined by its left and
right endpoints.  And, even more to the point, the number of vote
types is trivially at most linear in the size of the input, since each
vote type cast by at least one voter has to appear in the input.  So
the number of vote types certainly is not so large as to cause any
challenges (about having enough time to consider the vote types
occurring in $V'$) regarding remaining within polynomial time.

Now, at first, things might seem worrisome.  How can we decide which
votes to add, especially between ``incomparable'' vote types such as
(in our figure) the vote type with multiplicity 2 and the vote type
with multiplicity 7?  Those two vote types each help $p$ and some
other candidates, but the sets of other candidates helped are
incomparable---neither is a subset of the other.

We handle this worry by giving a ``smart greedy'' algorithm that only
makes moves that are clearly at least as safe as any other move, e.g.,
that will never close down the last path to success if any path to
success exists.  

Let us look again at the figure.  Note that although there are many
candidates, we can ignore all candidates except $p$ herself and the
candidates that are shaded---$l_6$, $l_3$, $l_2$, $r_3$, $r_6$, and
$r_{10}$.  Why?  Well, in our algorithm, we will never add any vote from
$V'$ that does not approve of $p$.  In fact, let us from now on
consider $V'$ to be redefined to have removed from it every vote that
does not approve of $p$.  (So in the figure, the $V'$ vote types 
shown
with multiplicity 5 and 9 go, as do any votes inside ``etc.''~that
do not approve of $p$.)
Now, since we'll never add a vote from $V'$
that does not approve of $p$
(indeed, we redefined $V'$ to drop such votes), 
and since each added vote will be an
interval that includes $p$, notice that, for example, if we add some
votes from $V'$ that cause $p$ to have (in total, between the original
votes from $V$ and added votes from $V'$) strictly more approvals than
$r_3$, then $p$ necessarily will have strictly more approvals than
$r_1$ and $r_2$, since each will have received from the added votes
from $V'$ no more additional approvals than $p$ received from those
added votes.  By the same reasoning, if by adding votes from 
(our redefined) $V'$ we
ensure that $p$ has strictly more approvals than $r_6$, clearly $p$
will have strictly more 
approvals than $r_4$ and $r_5$.\footnote{\singlespacing{}The 
  reader may worry that this ``by the same reasoning'' is not a valid
  claim, since one may worry that since $r_5$ starts above $p$ in
  score, we may by adding votes that approve of both $p$ and $r_5$
  (but not $r_6$) end up with $p$ having strictly more approvals than
  $r_6$ and yet with $r_5$ having more approvals than $p$, contrary to
  what our sentence above is asserting.
  The reason this reasonable worry is not actually a problem is that
  we will (as will be made clearer later in the proof)
  use a particular step-by-step process.  The process we follow will,
  before even thinking of adding any votes approving $r_5$, first have
  (if there is any path to ensuring this; if not, control is
  impossible) ensured \emph{via adding votes that give approval to $p$
    but not to $r_3$ (and thus not to $r_5$)} that $p$ has more
  approvals than $r_3$.  But note that after doing that, our ``by the
  same reasoning'' does cleanly apply to the case of $r_4$, $r_5$, and
  $r_6$.  This point is not identical to, but rather works
  hand-in-hand with, the point of the paragraph following the current
  paragraph.}
And the same
argument applies to the case of $r_{10}$ and, in mirror image, the
shaded left-side bars.  So, these are the only ``dangerous'' 
contenders that we have to worry about regarding $p$.  
This observation will guide our choice of which 
votes to add and which candidates to ignore.

Before going on, let us mention why we in the above went step-by-step.
That is, one might ask why we did not simply say that if $p$ by adding
votes from 
(our redefined) 
$V'$ could be made to defeat $r_{10}$ then it necessarily
would defeat all of $r_1$ through $r_9$.  But that claim in fact
does not hold.
Consider the case of adding 6 of the vote type having
multiplicity~7 in the figure.  
Doing so boosts $p$ to 8 approvals, passing $r_{10}$'s 7
approvals.  But in the process, $r_3$ rode the wave and was boosted to
9, and so $p$ does not beat $r_3$. That is, the argument of the
previous paragraph worked because the candidates we argued we could
ignore were strictly less approved, currently, than the ``dangerous''
voter we left in that outflanked them.  Just to be utterly specific,
the first dangerous candidate to the right of $p$ will be the 
leftmost 
candidate to the right of $p$ that from $V$ gets at least as many
approvals as $p$ (if we were in the nonunique-winner model, we would
have said ``strictly more approvals than $p$'').  In the figure, that
is $r_3$.  And the next dangerous candidate is whichever candidate is
the next candidate to the right of that candidate that from $V$ has
strictly more approvals than that candidate.  In our figure, that is
$r_6$.  And so on.  And analogously on the left.

So, now that we've set out who the dangerous rivals are, let us see if
there is a choice of control action that can let us beat them. In this
proof, we'll show just how to beat the very first of them, and then
will mention how to continue the process to (attempt to) defeat the
rest of them.  Crucially, the path we'll take to defeat our first
rival will be carefully chosen so that it is the safest possible
path---that is, if there is any overall successful control action,
then there will be one consistent with the actions we take to beat the
first rival (and the iterative repeating of the scheme we outline in
this proof will in fact then find such actions).

So let us focus on the first dangerous rival.  In our figure, that
would be $r_3$.  (Note: If there are no dangerous rivals, then $p$ is 
already a unique winner, and we're done.  If the only dangerous rivals we have
are on the left, then mirror image the universe, i.e., reverse $L$,
and now all our dangerous
rivals are on the right.)  Now, to become the unique
winner, $p$ must strictly beat $r_3$.  But note that among all the
votes in $V'$, the only ones that can possibly make that happen are
(keeping in mind that we consider $V'$ to now have in it only votes
that approve of $p$) \emph{those votes in $V'$ whose right endpoint
falls in the half-open interval $[p,r_3)$}.  
Call this set $B$.
These are the only votes
in $V'$ that let $p$ gain approvals relative to $r_3$.  So, in any
control action at all 
that causes $p$ to strictly beat (and we mention in passing that if we
were in the nonunique-winner model, we'd here say ``tie or beat'')
$r_3$, we must put enough votes from that collection, $B$, in to make
that happen.  So our only question is which of these votes to add (if
there even are enough of them to add to lead to success).  We do so as
follows.  We'll add votes from among these \emph{starting first with
the rightmost left endpoint}.  This is a perfectly safe strategy, as
that choice helps the same or strictly fewer ``left'' dangerous
candidates than would choosing a vote with a left endpoint that goes
more to the left.  As to the worry that by within $B$ looking only at
the left endpoint we may favor some \emph{right-side} candidates that
we might have otherwise slighted, that is so, but it is not a worry.
We indeed are treating all endpoints between $p$ and just before $r_3$
as a big blurred class, but the differences among those endpoints
affect just the 
(necessarily nondangerous)
candidates in-between, namely, in our figure,
candidates $r_1$ and $r_2$, but we already argued that by handling
$r_3$ we'll handle both of those, so giving extra approvals to one or
both of those (that we could have avoided by making another choice
within $B$) is not an issue.  In short we've intentionally smeared our
right-end finickiness because it doesn't matter, but left-end details
matter a lot.  So, in our figure, we need $p$ to gain two approvals
relative 
to $r_3$.  Using the scheme/ordering just mentioned, we'll among the
votes in $B$ keep eating through votes until we either achieve this or
until we've hit the control problem's limit on the total number of
votes we are allowed to add from $V'$.  In this case, in our figure's
example, the votes belonging to $B$ (let us assume that the
``etc.''~members of $V'$ contain no members of $B$) are those votes
contained inside the types $[l_4,p]$, $[l_3, r_1]$, and $[l_1,r_2]$.
The first vote type to draw on among these is (not withstanding that
doing so helps $r_2$, unlike the case for the other two members of
$B$) $[l_1,r_2]$, as it has the rightmost left endpoint.  That vote
type is of multiplicity 1, so we take that one vote.  The next vote
type to draw on is $[l_3, r_1]$.  Even though that vote type is of
multiplicity 2, $p$ only needs one more approval to beat $r_3$, so we
add to our election just one of the two of the votes of type $[l_3,
r_1]$.

And so we have defeated our first dangerous rival, and in the safest
possible way---a way that preserves a path to success if any exists.

The completion of the algorithm is basically to do this iteratively as
long as candidates remain that (relative to the current state of the
votes) are dangerous.  So, after we remove a rival, we can consider
all the added votes as if they are part of $V$ now, and can redraw the
figure, and can move forward, and do all the above against the next
dangerous candidate on the right (of course, decreasing the limit on
how many votes we are allowed to add by the number we just added), or
if we run out of candidates on the right, we can mirror image $L$ and
start eating through any remaining dangerous
candidates.  (Regarding who the
dangerous candidates are, the recomputation just mentioned is natural.
But we mention in passing that the set of right-side dangerous
candidates we'll deal with in the algorithm is in fact the same as the
set initially identified---the picture updating won't change it.
However, while handling those candidates, it is possible that some of
the additional approvals given to $p$ will help it defeat some of the
original left-side rivals, as a free side effect.)  If we achieve our
control goal before burning through the $k$ 
allowed additional votes from $V'$ (recall that $k$ is
part of the problem's input)
then we have
achieved our control goal, and otherwise it holds that our control goal
cannot be achieved.

Let us now turn to the case of deleting voters.  Since this case can
be handled essentially symmetrically to the adding-voters case (except
that we now have a very different choice of which voters to focus on
and how to sort them in order to decide which to delete), we merely
highlight the differences between the two cases.

In the deleting-voters case, a picture showing our initial state would
show just the bar chart of Figure~\ref{f:lane2}, i.e., the number of
approvals from $V$ each candidate gets initially.  After deletion, the
deleted votes would show up below the bar chart, each again as an
interval corresponding to the candidates approved of by this voter.
However, since it would be pointless to delete a voter approving
of~$p$, all such intervals do not include~$p$.

Now, we try to make $p$ strictly beat the
next dangerous rival to the right, starting with $r_3$ in the example
shown in Figure~\ref{f:lane2} (ignoring, of course, the intervals
indicating the votes in $V'$ and their multiplicities, as there is no
$V'$ in the deleting-voters case).  To do so, we now consider all the
(remaining, i.e., not yet deleted in the current iteration of our
algorithm) votes with intervals whose left endpoint falls in the
half-open interval $(p,r_3]$ (in the first iteration in our example).
Now we sort these votes \emph{from the rightmost right endpoint to the
  leftmost right endpoint}.  As in the adding-voters case (except
there we added votes), we now delete (following the sorted order) just
enough votes so that $p$ has one more approval than~$r_3$. If that is
impossible, we cannot control.  But if it is possible, we've succeeded
in making $p$ defeat $r_3$ in the safest possible way.  Recomputing
the figure (where, in the redrawn picture, updating the number of the
candidates' approvals implies that $r_3$ no longer is a dangerous
candidate for~$p$), we try to keep defeating $p$'s remaining dangerous
rivals, and when we run out of dangerous candidates to the right
of~$p$, we again mirror image the universe by reversing~$L$ and try to
make $p$ defeat any remaining dangerous candidates.  Eventually, this
process ends when we either succeed in making $p$ the unique winner by
defeating all of $p$'s dangerous rivals, or we exceed the deletion
limit $k$ before reaching this goal.  All details not discussed here
are obvious from the adding-voters proof.~\end{proofs}

We turn from approval voting to plurality voting. Plurality voting's
constructive control complexity was studied in detail in the seminal
control paper of Bartholdi, Tovey, and
Trick~\cite{bar-tov-tri:j:control}, and the
destructive cases were added by~\cite{hem-hem-rot:j:destructive-control}
(see~\cite{fal-hem-hem-rot:j:llull}
for the limited adding of candidates cases).  For plurality, the
situation is close to backwards from approval.  Regarding all the
adding/deleting control cases defined in this paper, the four
voter cases are vulnerable but all six candidate cases are resistant.
However, our results here again are in keeping with our paper's theme:  All
six of these cases become vulnerable for single-peaked societies.

\begin{theorem}\label{t:six-plurality}
For the single-peaked case, plurality voting is vulnerable to constructive
and destructive control by adding candidates, by adding unlimited
candidates, and by deleting candidates, and these results hold in both
the unique-winner model and the nonunique-winner 
model.\footnote{\singlespacing{}This result holds in our settled 
model in which the linear order specifying
the society's order on the candidates is part of the input, and also holds
in the model
in which the linear order is not part of the input but rather 
the question is whether there exists any valid 
linear order relative to which there is a way of 
achieving our control goal.}
\end{theorem}

Our proofs of the different parts of this theorem vary greatly in approach.
One approach that is particularly useful is dynamic programming.
We defer the proof of Theorem~\ref{t:six-plurality} to
Section~\ref{sec:proof-of-theorem-six-plurality}.

For two very important voting systems---plurality and approval---we have seen
that in every single adding/deleting case where they are known to have
NP-hardness complexity shields, the complexity shield evaporates for societies
with single-peaked preferences.
In the coming manipulation section we will also see a number of cases where
NP-hardness shields melt away for single-peaked societies, but we will also
see some
cases---and earlier such a case was found by
Walsh~\cite{wal:c:uncertainty-in-preference-elicitation-aggregation}---where
existing NP-hardness shields remain in place even if one adds the
restriction to a single-peaked society.

The previous paragraph raises the following natural question: Given that
restricting to single-peaked preferences can sometimes remove complexity
shields, e.g., Theorem~\ref{t:approval-control}, and can sometimes leave
them in place, e.g., Theorem~\ref{t:stays-hard}, can restricting to
single-peaked
preferences ever erect a complexity shield? It would be very tempting to
hastily state that such behavior is impossible.  After all, the
single-peaked case if anything thins out the flood of possible
control-actions/manipulations. But we're talking about complexity here, and
fewer options doesn't always mean a less complex problem.  In particular,
one of those manipulation/control-action options that single-peakedness
takes off the table might have been a single, sure-fire path to victory
under some 
election system.
In fact, using precisely that approach, and a few other tricks, we have
built an artificial election system for which unweighted constructive
manipulation by size-3 coalitions is in P in the general case but is
NP-complete for the single-peaked case. The system's votes are approval
vectors.  The system is highly unnatural,
highly unsatisfying, and would never be considered for real-world use.
However, its goal is just to show that (in a stomach-turningly contrived
way) restricting to single-peakedness in concept can, perhaps surprisingly,
raise complexity. We conjecture that that strange behavior will never be
seen for any existing, natural election system.  We summarize this
paragraph in the following (manipulation) theorem.

\begin{theorem}\label{t:erect-shield}
There exists an election system $\mathcal{E}$, whose votes are
approval vectors, for which constructive size-3-coalition unweighted
manipulation is in $\p$ for the general case but is $\np$-complete in
the single-peaked model.\footnote{\singlespacing{}This
result holds both in our settled model where $L$, the society's ordering of
the candidates, is part of the input and in the model where $L$
is not given and the question is whether there is any societal ordering of
the candidates that allows the manipulators to succeed.}
\end{theorem}

\begin{proofs}
Let $C = \{c_1, \ldots , c_m\}$ be a set of candidates that is
lexicographically ordered (i.e., $c_1$ is the smallest candidate in
this ordering; note that this lexicographical ordering is not related
to the society's ordering in single-peaked electorates) 
and let $V$ be a collection of approval vectors
over~$C$.  Given $C$ and $V$, election system $\mathcal{E}$ works as
follows.  If $\|C\| \leq 2$, all candidates win.  Otherwise (i.e., if
$\|C\| \geq 3$) we have two possibilities:
\begin{enumerate}
\item If there exists no permutation making the given votes
single-peaked (which, by Theorem~\ref{t:embed}, can be tested in
polynomial time) then all candidates win.\footnote{\singlespacing{}
  This feature of the system allows a manipulating coalition to in the
  general (single-peakedness not required) case cast votes precluding
  single-peakedness, an attack the single-peaked case does not allow.}
\item
Otherwise (i.e., if there is some permutation witnessing single-peakedness of the votes),\footnote{\singlespacing{}For the single-peaked case,
  our model also allows the election system $\mathcal{E}$ to directly
  access the input permutation relative to which $V$ is single-peaked.
  For ``real'' election systems this would be irrelevant but regarding
  building an artificial counterexample system, such as $\mathcal{E}$,
  this may be important, although Theorem~\ref{t:erect-shield} holds
  in either case.  Assuming that the election system knows the
  society's permutation witnessing single-peakedness, as our model
  does, matches the intuition that the societal ordering of candidates
  is open information.  Also, this assumption is fairer considering
  that our manipulation algorithms do get direct access to the input
  permutation of society, and our model is that that permutation is
  fixed (in candidate/voter addition cases it already must handle
  \emph{all} candidates and \emph{all} voters, and in candidate/voter
  deletion cases it must stick with the permutation even if after
  deletion other permutations are consistent with the remaining
  candidates/voters).  The permutation reflects ``society'' and does
  not change from what the input says it is.}
if the lexicographically smallest candidate, $c_1$, when viewed as
encoding a boolean formula encodes a satisfying one and the
lexicographically smallest approval vector is all zeros, then all
candidates win, else all candidates lose.
\end{enumerate}
For the general case, the constructive size-3-coalition unweighted
manipulation problem is in $\p$: Always have the three manipulators
cast the following votes (with respect to the lexicographic ordering
of $C = \{c_1, \ldots , c_m\}$):
\begin{eqnarray*}
(1, 1, 0, 0, \ldots , 0) \\
(0, 1, 1, 0, \ldots , 0) \\
(1, 0, 1, 0, \ldots , 0)
\end{eqnarray*}
ensuring that the electorate is not single-peaked, and so all
candidates win, including the distinguished candidate.

For the single-peaked case, however, the constructive size-3-coalition
unweighted manipulation problem is $\np$-complete.  The problem is
clearly in~$\np$, and to show $\np$-hardness we reduce the well-known
$\np$-complete boolean formula satisfiability problem to it.  Given a
boolean formula
$\varphi$ (encoded as a string in $\{0,1\}^*$), map it to the instance
$(C,V,\varphi)$ of the constructive size-3-coalition unweighted
manipulation problem, where $C = \{\varphi, 1^{|\varphi|+1},
1^{|\varphi|+2}\}$, $V$ consists of one nonmanipulator who approves of
all candidates and three manipulators, and where $\varphi$ is the
distinguished candidate the manipulators want to make win.  We claim
that they can succeed in doing so exactly if $\varphi$ is satisfiable.
Indeed, if $\varphi$ is satisfiable then $\varphi$ can be made a
winner by one of the manipulators casting the vote $(0,0,0)$.  On the
other hand, if $\varphi$ is not satisfiable (and the electorate is
restricted to being single-peaked) then $\varphi$ cannot win, no
matter which votes the manipulators cast.~\end{proofs}

\subsection{The Deferred Proof of Theorem~\ref{t:six-plurality}}
\label{sec:proof-of-theorem-six-plurality}

In this section we provide, as a series of lemmas that cover
appropriate cases, our proof of Theorem~\ref{t:six-plurality}.  One of
the reasons why candidate-control problems for plurality are hard in
the general case (i.e., when the voters are not restricted to be
single-peaked) is that then adding or deleting even a single candidate
can possibly affect the scores of all the other candidates. However,
in the single-peaked case adding or deleting a candidate can affect at
most two other candidates.  This observation, formalized in the next
lemma, is at the heart of the vulnerability proofs of this section.

\begin{lemma}\label{lem:scoringlemma}
  Let $(C,V)$ be an election where $C = \{c_1, \ldots, c_m\}$ is a set
  of candidates, $V$
  is a collection of voters
  whose preferences are single-peaked with respect to a linear order
  $L$, and where $c_1\, L\, c_2\, L \,\cdots\, L\, c_m$. Within plurality,
  if $m \geq 2$ then
  \begin{enumerate}
  \item $\score_{(C,V)}(c_1) = \score_{(\{c_1,c_2\},V)}(c_1)$,
  \item for each $i$, $2 \leq i \leq m-1$, $\score_{(C,V)}(c_i) =
    \score_{(\{c_{i-1},c_{i},c_{i+1}\},V)}(c_i)$, and
  \item $\score_{(C,V)}(c_m) = \score_{(\{c_{m-1},c_m\},V)}(c_m)$.
  \end{enumerate}
\end{lemma}
\begin{proofs}
  Let $(C,V)$, $L$, and $m$ be as in the statement of the lemma.  We
  prove the second case, $2 \leq i \leq m-1$ (the proofs of the
  boundary cases are analogous). Fix an integer $i$, $2 \leq
  i \leq m-1$.  Each voter $v$ in $C$ that ranks $c_i$ as her top
  choice, clearly, still prefers $c_i$ to both $c_{i-1}$ and
  $c_{i+1}$ when limited to the choice between $c_i$, $c_{i-1}$, and
  $c_{i+1}$. On the other hand, consider a voter $v$ in $V$ who ranks
  some candidate $c_j$, $c_i \neq c_j$, as her most preferred
  candidate. If $j \leq i-1$ then, due to the fact that preferences
  are single-peaked (or, $j = i-1$), this voter prefers $c_{i-1}$ to
  $c_i$. If $j \geq i+1$ then, for the same reason, this voter prefers
  $c_{i+1}$ to $c_i$. In either case, this voter does not give her
  point to~$c_i$. As a result, $\score_{(C,V)}(c_i) =
  \score_{(\{c_{i-1},c_{i},c_{i+1}\},V)}(c_i)$.~\end{proofs}

Using Lemma~\ref{lem:scoringlemma}, we show that in the
single-peaked case plurality is vulnerable to constructive control by
adding candidates.

\begin{lemma}
\label{lem:six-plurality-constructive-adding-candidates-unique}
For the single-peaked case, plurality voting is vulnerable to
constructive control by adding candidates both in the unique-winner model
and in the nonunique-winner model.
\end{lemma}

\sproof
We give a polynomial-time algorithm for the single-peaked variant of
the constructive control by adding candidates problem.  We focus on
the unique-winner case, but the proof can easily be adapted to the
nonunique-winner model and we will be pinpointing the necessary
changes throughout the proof.
Our input is an election $(C \cup A,V)$---where $C = \{p,c_1,
\ldots, c_{m'}\}$ and $A = \{a_1, \ldots, a_{m''}\}$---a candidate $p$ in
$C$, a nonnegative integer $k$, and a linear order $L$ over $C\cup
A$.\footnote{\singlespacing{}If we are in the model in which no linear
societal order is given simply use, for example, 
the algorithm of Escoffier, Lang, and
{\"O}zt{\"u}rk~\cite{esc-lan-ozt:c:single-peaked-consistency} to find
one relative to which our input is single-peaked.  If none exists,
state that and terminate.  If the algorithm
of~\cite{esc-lan-ozt:c:single-peaked-consistency} gives us some valid
order, it is not hard to see that the algorithm we are about to describe
in this proof will find a successful control action exactly if there is
any valid societal linear order relative to which a successful control
action exists.  (In the remaining proofs of
Section~\ref{sec:proof-of-theorem-six-plurality}, we'll handle only the
case that the societal order witnessing single-peakedness is given
in our polynomial-time algorithms showing vulnerability claims.  Note,
however, that this footnote's argument handles the case of such a
societal order not being given analogously in each such case.)
\label{foo:linear-order}}
After checking that the voters in $V$ are single-peaked with
respect to~$L$,
we ask if it is possible
to find a subset $A'$ of $A$ such that (a)~$\|A'\| \leq k$ and (b)~for
each $c \in C \cup A'$, $c \neq p$, it holds that $\score_{(C \cup
  A',V)}(c) < \score_{(C\cup A',V)}(p)$ ($\score_{(C \cup A',V)}(c)
\leq \score_{(C\cup A',V)}(p)$, in the nonunique-winner case).
Let $D = C \cup A$ and let us rename the candidates so that $D =
\{d_1, \ldots, d_m\}$, where $m = 1 + m' + m''$, and $d_1\, L\, d_2\, L
\cdots\, L\, d_m$. Without loss of generality we can assume that
$p$ is neither $d_1$ nor $d_m$. If this were not the case, we
could add to $C$ two dummy candidates, $d'$ and
$d''$, that each voter ranks as the least desirable ones and such
that $d'\, L\, d_1\, L\, \cdots\, L\,d_{m}\, L\,d''$.

Before we describe the algorithm formally, let us explain it
intuitively.  Let us assume that our instance is a ``yes'' instance
and that $A'$ is a subset of $A$ witnessing this fact. There must be two
candidates in $C \cup A'$, $d_\ell$ and $d_r$, that are direct
neighbors of $p$ with respect to $L$ restricted to $C \cup A'$ (i.e.,
$d_\ell\, L\, p\, L\, d_r$ and there is no $d_i \in C \cup A'$ such
that $d_i \neq p$ and $d_\ell\, L\, d_i\, L\, d_r$).  By
Lemma~\ref{lem:scoringlemma}, the score of $p$
is
``fixed'' by $d_\ell$ and $d_r$ in the sense that $\score_{(C\cup
  A',V)}(p) = \score_{(\{d_\ell,p,d_r\},V)}(p)$. Also, given such
$d_\ell$ and $d_r$ we can view $C \cup A'$ as partitioned into the
left part, $C_\ell \cup A'_\ell$, containing $d_\ell$ and the
candidates preceding her, the right part, $C_r \cup A'_r$,
containing $d_r$ and the candidates succeeding her, and the
middle part, containing $p$.  The candidates in the left part and the
candidates in the right part all have scores lower than~$p$.

The idea of the algorithm is to try all possible pairs of candidates
$d_\ell$ and $d_r$ that can be direct neighbors of $p$ and to test,
for each such pair, if it is possible to add candidates so that (a)
each of the candidates in the left part has a score lower than~$p$, (b) each
of the candidates in the right part has a score lower than~$p$, and (c)
the number of candidates added is at most~$k$. This approach is
formalized in the algorithm below.

\begin{enumerate}
\item For each pair of candidates, $d_\ell$ and $d_r$, such that
  \begin{enumerate}
  \item $d_\ell\, L\, p\, L\, d_r$, and
  \item there is no candidate $c_i \in C$ such that $d_\ell\, L\,
    c_i\, L\, d_r$,
  \end{enumerate}
  execute Steps~\ref{step2} through~\ref{step5} below. After trying all
  pairs $d_\ell, d_r$ without accepting, reject.

\item\label{step2} 
  \begin{enumerate}
  \item\label{step2a} Set $b' = \score_{(\{d_\ell,p,d_r\},V)}(p) - 1$.
  \item Set $C_\ell = \{ c_i \mid c_i\, L\, d_\ell\} \cup \{d_\ell\}$ and
    $C_r = \{ c_i \mid d_r\, L\, c_i \} \cup 
    \{d_r\}$.\footnote{\singlespacing{}Yes,
      we put $d_\ell$ in $C_\ell$ and $d_r$ in $C_r$ even if either of
      them belongs to $A$.}
  \item Set $A_\ell = \{ a_i \mid a_i\, L\, d_\ell\}$ and
            $A_r = \{ a_i \mid d_r\, L\, a_i \}$.
  \item $V_\ell$ is the collection of those voters in $V$ who, among the
    candidates in $C_\ell \cup C_r \cup A_\ell \cup A_r \cup \{p\}$, rank
    first either $d_\ell$ or a candidate $d_i$ such that $d_i\, L\,
    d_\ell$; $V_r$ is the collection of those voters in $V$ who,
    among the candidates in $C_\ell \cup C_r \cup A_\ell \cup A_r \cup \{p\}$,
    rank first either $d_r$ or a candidate $d_i$ such that $d_r\, L\,
    d_i$.\footnote{\singlespacing{}Note that 
      if $B_\ell$ is a subset of $A_\ell$ and
      $B_r$ is a subset of $A_r$ then in election $(C \cup B_\ell \cup
      B_r \cup \{d_\ell,d_r\},V)$ it holds that (a)~each voter in
      $V_\ell$ ranks first some candidate in $C_\ell \cup B_\ell \cup
      \{d_\ell\}$, (b)~each voter in $V_r$ ranks first some candidate
      in $C_r \cup B_r \cup \{d_r\}$, and (c)~all the remaining voters 
      rank $p$ first.}
 \end{enumerate}

\item\label{stepL} Find a smallest (with respect to cardinality)
  set $B_\ell$, $B_\ell \subseteq
  A_\ell$, such that for each $c \in C_\ell \cup B_\ell$ it holds that
  $\score_{(C_\ell \cup B_\ell,V_\ell)}(c) \leq b'$.  If such a set
  $B_\ell$ does not exist then discard the current pair $(d_\ell,d_r)$
  and try the next one.

\item\label{stepR} Find a smallest (with respect to cardinality)
  set $B_r$, $B_r \subseteq
  A_r$, such that for each $c \in C_r \cup B_r$ it holds that
  $\score_{(C_r \cup B_r,V_r)}(c) \leq b'$.  If such a set
  $B_r$ does not exist then discard the current pair $(d_\ell,d_r)$
  and try the next one.

\item\label{step5} If $\|B_\ell\| + \|B_r\| + \|A \cap
  \{d_\ell,d_r\}\| \leq k$ then accept (the set $A'$ that we seek is
  $B_\ell \cup B_r \cup (\{d_\ell,d_r\} \cap A)$).
\end{enumerate}

It is easy to see that the above algorithm is correct and that it also
works in the nonunique-winner model if we replace Step~\ref{step2a}
with $b' = \score_{(\{d_\ell,p,d_r\},V)}(p)$.  We will now show that
the algorithm runs in polynomial time.  There are at most
quadratically many (with respect to $m = \|D\|$) pairs $(d_\ell, d_r)$ to
try, so each of the above steps will be executed at most polynomially
often. It remains to show that Steps~\ref{stepL} and~\ref{stepR} can
each be executed in polynomial time.  To this end, we define the
following functional problem.

\begin{definition}
  In the DemoteByAddingCandidates problem we are given an election $(C
  \cup A,V)$, where the voters in $V$ are single-peaked with respect
  to a given order $L$, and a nonnegative integer $b$.  We ask what
  the size is of a smallest set $A' \subseteq A$ such that for each
  candidate $c \in C \cup A'$ it holds that $\score_{(C \cup A',V)}(c) \leq
  b$. If such a set $A'$ does not exist then, by convention, the
  answer is $\infty$.
\end{definition}

Computing $B_\ell$ ($B_r$) in Step~\ref{stepL} (in
Step~\ref{stepR}) simply requires solving an instance of
DemoteByAddingCandidates with election $(C_\ell \cup A_\ell,V_\ell)$
(election $(C_r \cup A_r,V_r)$), order $L$ (restricted to an
appropriate subset of candidates), and integer $b = b'$.  Fortunately,
DemoteByAddingCandidates is easy to solve.

\begin{lemma}\label{thm:dbac}
  DemoteByAddingCandidates is computable in polynomial time.
\end{lemma}
\sproof %
  We give a polynomial-time algorithm for the
  DemoteByAddingCandidates problem. Our input consists of an election
  $(C \cup A, V)$, a nonnegative integer $b$, and a linear order $L$
  over $C \cup A$, where $C = \{c_1, \ldots, c_{m'}\}$, $A = \{a_1,
  \ldots, a_{m''}\}$, and voters in $V$ have single-peaked preferences
  with respect to the 
  order $L$.\footnote{\singlespacing{}Keep in mind that within the proof
    of this lemma, $C$, $A$, $V$, $L$, and $b$ refer to the input to
    the DemoteByAddingCandidates problem, not to the input to the
    Control by Adding Candidates problem.}

  Let $D = C \cup A$ and let us rename the candidates so that $D =
  \{d_1, \ldots, d_m\}$, where $m = m' + m''$ and $d_1\, L\, d_2\, L\,
  \cdots\, L\, d_m$. Without loss of generality, we can assume that
  $d_1 \in C$ and that $m \geq 3$.\footnote{\singlespacing{}If
    $m$ were less than~$3$,
    we could add to $C$ an appropriate number of dummy candidates, each
    ranked by all the voters as less desirable than any of the
    candidates in $C \cup A$.  Extending $L$ so that for each added
    dummy candidate $d$ it holds that $d\, L\, d_1$ and the voters
    are single-peaked with respect to this extended $L$ is easy.}
  Let $i$, $j$, and $k$ be three
  positive integers such that $1 \leq i < j < k \leq m$. 
  We define
\begin{enumerate}
\item $s(i,j,k) = \score_{(\{d_i,d_j,d_k\},V)}(d_j)$,
\item $s(0,j,k) = \score_{(\{d_j,d_k\},V)}(d_j)$, and
\item $s(i,j,m+1) = \score_{(\{d_i,d_j\},V)}(d_j)$.
\end{enumerate}
  Note that, by Lemma~\ref{lem:scoringlemma}, if $B$ is a subset of $A$
  such that (a)~$\{d_i,d_j,d_k\} \subseteq C \cup B$ and (b)~there is
  no candidate $d_l$ ($l \neq j$) such that $d_i\, L\, d_l\, L\, d_k$,
  then $s(i,j,k) = \score_{(C\cup B, V)}(d_j)$.

  Let $i$ and $j$ be two positive integers such that $1 \leq i < j
  \leq m$. By $A(i,j)$ we mean a family of subsets of $A$ such that a
  set $B$, $B \subseteq A$, belongs to $A(i,j)$ if and only if (a)
  $d_i \in C \cup B$, (b) $d_j \in C \cup B$, and (c) $(C \cup B) \cap
  \{d_{i+1}, \ldots, d_{j-1}\} = \emptyset$.

  We now define a key value for our algorithm. Let $i$ and $j$ be
  two positive integers such that $1 \leq i < j \leq m$. We 
  define
  \[
    f(i,j) = \min_{B \in A(i,j)}\left\{ \|B\| \mid (\forall d_\ell \in
    C\cup B, 1 \leq \ell \leq i)[ \score_{(C \cup B,V)}(c_\ell) \leq b ]
    \right\}.
  \]
  We adopt the convention that
  $\min \emptyset$ is 
  $\infty$. 
  It is easy to verify that the
  algorithm's output should be
  \[
  \min_{1 \leq i < j \leq m} \left\{ f(i,j) \mid \{d_{j+1}, \ldots,
    d_m\} \subseteq A \land s(i,j,m+1) \leq b \right\}.
  \]
  It remains to show that we can compute the 
  values $f(i,j)$, $1 \leq i < j \leq m$,
  in polynomial time. The rest of the proof is devoted to this task. 

  We compute the
  values $f(i,j)$ using dynamic programming. Let $j$ be a positive
  integer $1 < j \leq m$. It is easy to see that 
  \[
  f(1,j) = \left\{ \begin{array}{ll}
    \infty & \mbox{if there is some $d_i \in C$ such that $1 < i < j$,}    \\
    \infty & \mbox{if the above does not hold but $s(0,1,j) > b$,}    \\
    \| A \cap \{d_1, d_j\}\| & \mbox{otherwise.} 
    \end{array}\right.
  \]
  Given these 
  initial values,\footnote{\singlespacing{}Note that here it is important
    that $d_1 \in C$. If this were not the case, we would also have to
    directly compute the values $f(i,j)$ for some $i,j$ where $1 < i <
    j$.}  we can compute arbitrary values $f(j,k)$, $1 < j < k \leq
  m$, using the following facts. Let us fix $j$ and $k$ such that $1 <
  j < k \leq m$.  If there is a $d_\ell \in C$ such that $j < \ell <
  k$ then $f(j,k) = \infty$ (because $A(j,k)$ is empty). Otherwise
  \[
   f(j,k) = \min_{1 \leq i < j}\{ f(i,j) + \chi_A(d_k) \mid s(i,j,k) \leq b\},
  \]
  where $\chi_A$ is the characteristic function of $A$ (i.e.,
  $\chi_A(d_k)$ is $1$ if $d_k \in A$, and is $0$ otherwise).  It is
  easy to see that using standard dynamic-programming techniques and
  the above equations we can compute $f(i,j)$ for arbitrary
  positive integers $i,j$, $1 \leq i < j \leq m$, in polynomial
  time.~\eproofof{Lemma~\ref{thm:dbac}}

This completes the proof that for the case of single-peaked voters,
plurality is vulnerable to constructive control by adding candidates.~\eproofof{Lemma~\ref{lem:six-plurality-constructive-adding-candidates-unique}}

As a corollary to 
Lemma~\ref{lem:six-plurality-constructive-adding-candidates-unique},
we obtain the corresponding result for the case of adding an unlimited
number of candidates, because for
each unlimited
case it is sufficient to use the algorithm for the corresponding
limited case with the limit on the number of candidates that can
be added set to equal the number of spoiler candidates.

\begin{corollary}
\label{cor:six-plurality-constructive-adding-candidates-unique}
For the single-peaked case, plurality voting is vulnerable to
constructive control by adding an unlimited number of candidates in
both the unique-winner model and the nonunique-winner model.
\end{corollary}

In the following proofs (both in the destructive adding-candidates
cases and in the deleting-candidates cases) in addition to
Lemma~\ref{lem:scoringlemma} we will need one more tool, namely, the
notion of a neighborhood of a candidate $c$ in an order
$L$ witnessing single-peakedness of an electorate.

\begin{definition}
Let $(C,V)$ be an
election and let $L$ be an order such that the voters in $V$ are
single-peaked with respect to $L$.  Let $c$ be a candidate in $C$. We
rename the candidates in $C$ such that $C = \{b_{m'}, \ldots, b_2,
b_1, c, d_1, d_2, \ldots, d_{m''}\}$ and
$
  b_{m'} \, L \, \cdots \, L \, b_2 \, L \, b_1    \, L \,
      c \, L \, d_1    \, L \, d_2 \, L \, \cdots \, L \, d_{m''}.
$
For each two positive integers $i$ and $j$ we set $D_{ij}(C,c) =
\{b_1,\ldots, b_{\min(i,m')}\} \cup \{d_1, \ldots,
d_{\min(j,m'')}\}$. We define the direct neighborhood of a candidate
$c$ to be the set $D(C,c) = \{ D_{ij}(C,c) \mid 0 \leq i \leq m'
\mbox{ and } 
0 \leq j \leq m''\}$.\footnote{\singlespacing{}The notation
  of this definition, in essence, ties the order $L$ to the
  candidate set $C$. We will sometimes wish to speak of, e.g.,
  $D_{ij}(C',c)$ for some $C' \subseteq C$, $c \in C'$, and this
  notation allows us to naturally, implicitly speak of an order $L'$ induced by
  $C'$.}
\end{definition}

Clearly, given a single-peaked election $(C,V)$ and a candidate $c
\in C$, $\|D(C,c)\|$ is polynomial in~$\|C\|$.
We now turn to the destructive adding-candidates
cases in both the unique-winner model and the nonunique-winner
model. The proof is much simpler than in the constructive cases.

\begin{lemma}
\label{lem:six-plurality-destructive-adding-candidates-unique}
For the single-peaked case, plurality voting is vulnerable to
destructive control by adding candidates in both the unique-winner
model and the nonunique-winner model.
\end{lemma}

\begin{proofs}
Let $(C \cup A, V)$ be an election where $C = \{d, c_1, \ldots,
c_m\}$, $A = \{a_1, \ldots, a_{m'}\}$, and $V$ is a collection of
voters whose preference orders are single-peaked with respect to a
given order $L$. The candidates from
$C$ are already registered 
and the candidates from
$A$ can be added (i.e., $A$
is the spoiler candidate set). Let $k$ be some nonnegative integer. We
now give a polynomial-time algorithm that tests whether there exists a
set $A' \subseteq A$ such that (a)~$\|A'\| \leq k$ and (b)~$d$ is not
a unique winner of $(C \cup A',V)$.

The algorithm works as follows: For each
up-to-3-element subset $A''$ of~$A$,
check if $d$ is a unique winner of $(C \cup A'',V)$
and accept if
this is not the case (for the first such $A''$ found).
If we have not accepted for any such subset $A''$ then reject.
This algorithm clearly works in polynomial time. It remains to show
that it is correct.  If for each $A' \subseteq A$ it holds that $d$ is
a unique winner of $(C \cup A',V)$ then the above algorithm
(correctly) rejects. On the other hand, let us assume that for some
$A' \subseteq A$ it holds that $d$ is not a unique winner of $(C \cup
A',V)$ and let us fix one such $A'$.  We claim that there is an
up-to-size-3 subset $A''$ of $A$ such that $d$ is not a unique winner
of $(C \cup A'',V)$.

Let $B = D_{1,1}(C \cup A',d)$, that is, $B$ contains the direct
neighbors of $d$ (among the candidates in $C \cup A'$, with respect to~$L$).
By definition, $\|B\| \leq 2$, and, from
Lemma~\ref{lem:scoringlemma}, $\score_{(C\cup A',V)}(d) =
\score_{(B\cup\{d\},V)}(d)$. Since $d$ is not a unique winner of
$(C\cup A',V)$, there is some candidate $c \in C \cup A'$, $c \neq d$,
such that $\score_{(C \cup A',V)}(c) \geq \score_{(C\cup
  A',V)}(d)$. Define $A'' = (B \cup \{c\}) \cap A$. Clearly,
$\|A''\| \leq 3$ and, since $B$ is a ``radius-1'' neighborhood of
$d$, we have $\score_{(C \cup A'',V)}(d) =
\score_{(B\cup\{d\},V)}(d)$. On the other hand, since $C \cup A''
\subseteq C \cup A'$ and $c$ belongs to both sets, it holds that
$\score_{(C \cup A'',V)}(c) \geq \score_{(C \cup A',V)}(c)$.
It follows that in
election $(C \cup A'',V)$ candidate $c$ has a score
at least as high as that of $d$
and,
thus, $d$ is not a unique winner of $(C \cup A'',V)$.

It is easy to see that the same approach can be used to solve the
nonunique-winner case. Our algorithm now tries all up-to-3-element
subsets $A''$ of $A$ and accepts if and only if there is one such that
$d$ is not a winner of $(C \cup A'',V)$. The proof of correctness is
analogous to the unique-winner case.  This completes the
proof.~\end{proofs}

As
an easy corollary we obtain that in the single-peaked case
plurality is also vulnerable to destructive control by adding unlimited
candidates.

\begin{corollary}
  For the single-peaked case, plurality voting is vulnerable to
  destructive control by adding an unlimited number of candidates in
  both the unique-winner model and the nonunique-winner model.
\end{corollary}

We now turn to the constructive and destructive deleting-candidates
cases in both the unique-winner model and the nonunique-winner model.
The main idea of our algorithms for the deleting-candidates cases is
that it is sufficient to focus on deleting candidates
adjacent to the
preferred one (to increase her score) and then to delete those
remaining candidates that still defeat the preferred one.

\begin{lemma}
\label{lem:six-plurality-constructive-destructive-deleting-candidates-unique-nonunique}
For the single-peaked case, plurality voting is vulnerable to
constructive and destructive control by deleting candidates in both the
unique-winner model and the nonunique-winner model.
\end{lemma}

\begin{proofs}
  We give a polynomial-time algorithm for the single-peaked case
  of constructive control by deleting candidates for plurality.  We
  focus on the unique-winner model but it is easy to see how to modify
  our algorithm to work in the nonunique-winner model.

  Our algorithm's input contains an election $(C,V)$, a preferred
  candidate $p \in C$, a nonnegative integer $k$ (the number of
  candidates that we are allowed to delete), and an order $L$ such
  that the voters in $V$ are single-peaked with respect to~$L$.

  By Lemma~\ref{lem:scoringlemma}, we see that the only possible scores
  that $p$ can have after deleting some candidates are in the set $S =
  \{ \score_{(C-D,V)}(p) \mid D \in D(C,p) \}$ and that each of these
  scores can be obtained by deleting some subset $D \in D(C,p)$ of the
  candidates.
  In consequence, to
  ensure that $p$ is the unique winner we first need to delete some subset
  $D \in D(C,p)$ of candidates and then---since it is impossible to
  decrease a candidate's score via deleting other candidates---delete
  all those candidates that still have more points than $p$.  Of
  course, we do not know \emph{a priori}
  which set $D$ in $D(C,p)$ to delete, but since
  $\|D(C,p)\|$ is polynomially bounded in $\|C\|$, we can try all its
  members.

  Formally, our algorithm works as follows.

  \begin{enumerate}
  \item \label{step:plurality-constructive-destructive-deleting-1}
  For each $D \in D(C,p)$, execute
  Step~\ref{step:plurality-constructive-destructive-deleting-2}.
  If, after trying all $D$'s, we do not accept, reject.
  \item \label{step:plurality-constructive-destructive-deleting-2}
    \begin{enumerate}
    \item \label{dcacplurality2b} While there exists a 
      candidate~$c \in C-D$, $c \neq p$,
      such that $\score_{(C-D,V)}(c) \geq
      \score_{(C-D,V)}(p)$, add $c$ to~$D$.
    \item If $\|D\| \leq k$ then accept.
    \end{enumerate}
  \end{enumerate}
  Clearly, this algorithm works in polynomial time and, by the
  preceding discussion, it accepts if and only if it is possible to
  ensure $p$'s victory via control by deleting candidates.  For the
  nonunique-winner case, we simply need to change the inequality in
  Step~\ref{dcacplurality2b} from ``$\geq$'' to ``$>$.''

  In the destructive case we are given an election $(C,V)$, a despised
  candidate $d \in C$, a nonnegative integer $k$ (the number of
  candidates we can delete), and an order $L$ such that the voters in
  $V$ are single-peaked with respect to~$L$. (We assume that $d$ is a
  unique plurality winner of $(C,V)$; otherwise, we can immediately
  accept.) It is sufficient to find a candidate $c$ and a subset $D
  \subseteq C - \{d\}$ of candidates such that $\|D\| \leq k$ and
  $\score_{(C-D,V)}(c) \geq \score_{(C-D,V)}(d)$.  As in the
  constructive case, by Lemma~\ref{lem:scoringlemma}, it suffices to
  try each candidate $c \in C - \{d\}$ and each set $D \in D(C,c)$. That
  is, if there is a candidate $c \in C - \{d\}$ and a set $D \in D(C,c)$
  such that (a)~$\|D\| \leq k$, (b)~$d \notin D$, and
  (c)~$\score_{(C-D,V)}(c) \geq \score_{(C-D,V)}(d)$, then we accept.
  Otherwise we reject. Clearly, this algorithm works in polynomial
  time and, by Lemma~\ref{lem:scoringlemma} and the fact that deleting
  candidates does not decrease the remaining candidates' scores, is
  correct. The algorithm can be modified in an obvious way to work for the
  nonunique-winner model.~\end{proofs}

\section{Manipulation}\label{sec:manipulation}

In this section we study constructive coalition weighted manipulation.
Recall that in the single-peaked case the manipulators must cast votes
that are consistent with the linear ordering of the candidates (which is part
of the input in the single-peaked case) that defines the society's
single-peakedness. However,
all our ``single-peaked case is in P'' results in this section also hold
in a model in which the linear order of the candidates is not part of the
input and we will comment on these cases in our proofs below.

This paper's theme is that single-peakedness removes many NP-hardness
shields, and for manipulation we show that via
Theorem~\ref{t:fall-down} (and its implications within
Theorem~\ref{t:toodles-csl}) and Theorem~\ref{t:3-candidate-dichotomy}.
Note
that in the general case, the election systems of parts~\ref{part:borda-three}
and~\ref{part:veto} of Theorem~\ref{t:fall-down}
and the ``$k_1 \geq 2 \wedge k_0 \geq 1$'' cases of
part~\ref{part:majority-ones} of this theorem
are known to be NP-complete~\cite{hem-hem:j:dichotomy,pro-ros:j:juntas,con-lan-san:j:few-candidates},
and the remaining
part~\ref{part:majority-ones} 
cases are
easily seen to be in P~\cite{hem-hem:j:dichotomy}.)

\begin{theorem}\label{t:fall-down}
For the single-peaked case, the constructive coalition weighted manipulation
problem (in both the nonunique-winner model and the unique-winner model)
for each of the following election systems is
in~$\p$:
\begin{enumerate}
\item\label{part:borda-three}
The scoring protocol $\alpha = (2,1,0)$, i.e., 3-candidate Borda elections.
(This is a special case of Theorem~\ref{t:3-candidate-dichotomy}.)
\item\label{part:majority-ones} 
Each scoring protocol $\alpha = 
(\overbrace{1,\ldots,1}^{k_1},\overbrace{0,\ldots,0}^{k_0})$, 
$k_1 \geq k_0$.  (This includes a variety of $\ell$-veto and 
$\ell'$-approval protocols, e.g., the 3-veto cases for $m \geq 6$
candidates in Theorem~\ref{t:toodles-csl}.)
\item\label{part:veto} 
Veto.\footnote{\singlespacing{}All three of Theorem~\ref{t:fall-down}'s cases 
hold both in our settled model where $L$, the society's ordering of
the candidates, is part of the input and in the model where $L$
is not given and the question is whether there is any societal ordering of
the candidates that allows the manipulators to succeed.}
\end{enumerate}
\end{theorem}

\begin{proofs}
For each of the election systems stated in the theorem, we give a
polynomial-time algorithm solving the constructive coalition weighted
manipulation problem for the single-peaked case.  Each of these
algorithms takes as input a list $C$ of candidates, a list $S$ of
nonmanipulative voters each specified by a single-peaked preference
over $C$ (with respect to the given linear ordering $L$ on $C$ that
defines single-peakedness) and an integer weight, a list of the
weights of the voters in our manipulating coalition~$T$, and the
candidate $p \in C$ that the manipulators in $T$ seek to make a winner
(a unique winner).  Note that all manipulators in $T$ must also have
single-peaked preferences with respect to~$L$.

  \medskip
  \noindent
  {\bf The scoring protocol \boldmath{$\alpha = (2,1,0).$}}\quad
  We start with the
  nonunique-winner case. 
  Let $C = \{a,
  b, p\}$.  Without loss of generality, it is enough to consider the
  following two societal orderings~$L$.  (If $L$ is not part of the
  input, we simply try all societal orders.)

\begin{description}
\item[Case~1:] $a \,L\, p \,L\, b$.  Due to single-peakedness, only
  the following votes are allowed with respect to~$L$:
\begin{equation*}
p > a > b, \quad
p > b > a, \quad
a > p > b, \quad
b > p > a. 
\end{equation*}
Note that, regardless of the weights of the voters in~$S$, we have
$\score_{S}(p) \geq \score_{S}(a)$ or $\score_{S}(p) \geq
\score_{S}(b)$.  So, at most one of $a$ and $b$ beats~$p$,
say~$a$.  Set all votes in $T$ to $p > b > a$.

\item[Case~2:] $p \,L\, a \,L\, b$.  In this case, due to
  single-peakedness, only the following votes are allowed with respect
  to~$L$:
\begin{equation*}
p > a > b, \quad
a > p > b, \quad
a > b > p, \quad
b > a > p.
\end{equation*}
In this case, a best vote for all voters in $T$ is $p > a > b$.
\end{description}
In both cases, we accept if the manipulation action described above
was successful (i.e., made $p$ a winner in election $(C, S \cup T)$),
and we reject otherwise.

To handle the unique-winner case, note that Case~2 remains the same.
For Case~1, even if
$\score_{S}(b) = \score_{S}(p)$, setting all voters in $T$ to
$p > b > a$ will work and is still optimal.  In both cases, for
checking success of the manipulation action we now test whether it
makes $p$ the unique winner.

\medskip
\noindent
{\bf Scoring protocols \boldmath$\alpha = 
(\overbrace{1,\ldots,1}^{k_1},\overbrace{0,\ldots,0}^{k_0})$, $k_1 \geq k_0$.}\quad
Let $\alpha =
(\overbrace{1,\ldots,1}^{k_1},\overbrace{0,\ldots,0}^{k_0})$, $k_1\geq k_0$,
be a scoring protocol.
We will handle the cases $k_1 > k_0$ and $k_1 = k_0$
separately.
As above, we will present algorithms for both the nonunique-winner
model and the unique-winner model.

First, assume $k_1 > k_0$.  In any election with $m = k_0 + k_1$
candidates, if the voters are single-peaked with respect to some
linear 
order~$L$,\footnote{\singlespacing{}If $L$ is not part of the input, we simply
  try all possible societal orderings.  This
  can be done in polynomial time, since there are only a constant
  number of candidates.
\label{foo:try-all-orderings}}
then the middle candidate(s) (namely, the $\lceil \nicefrac{m}{2}
\rceil$nd candidate in $L$ if $m$ is odd, and the
$(\nicefrac{m}{2})$nd and the $(1+\nicefrac{m}{2})$nd candidate in $L$
if $m$ is even) will always be among the winners.
Thus, given an instance of the
constructive coalition weighted manipulation problem, our
distinguished candidate $p$ can be made a winner in $(C,S \cup T)$ if
and only if $p$ does not lose a point in~$S$.  So our algorithm checks
whether $p$ does not lose a point in~$S$ and accepts or rejects
accordingly.

For the unique-winner case within scoring protocol $\alpha =
(\overbrace{1,\ldots,1}^{k_1},\overbrace{0,\ldots,0}^{k_0})$, $k_1 >
k_0$, $p$ can be made a unique winner if and only if
\begin{enumerate}
\item $p$ does not lose a point in~$S$, and
\item for all candidates $c$ that are tied with $p$ in $S$, it is
  possible to make $c$ lose a point in $T$ (while keeping $T$
  single-peaked with respect to~$L$).
\end{enumerate}
Since the number of candidates is fixed, our algorithm can easily
check whether these two conditions hold and then accepts or rejects
accordingly.

Now, assume $k_1 = k_0$.  Given an instance of the constructive
coalition weighted manipulation problem, it is enough to consider a
linear order $L$ (which witnesses single-peakedness; also, recall
Footnote~\ref{foo:try-all-orderings}) that ranks at least $k = k_0 =
k_1$ candidates before $p$ and fewer than $k$ candidates after~$p$:
$c_1 \,L\, c_2 \,L\, \cdots \,L\, c_{\ell} \,L\, p \,L\, c_{\ell+2}
\,L\, \cdots \,L\, c_{2k}$ with $\ell \geq k$.  For every candidate
$c_i$, $\ell + 2 \leq i \leq 2k$, to the right of~$p$, we have
$\score_{S}(p) \geq \score_{S}(c_i)$, and for every candidate $c_j$,
$k+1 \leq j \leq \ell$, to the left of $p$ and to the right of the
middle, it holds that if $p$ gets a point then so does~$c_j$.  It
follows that $p$ can be made a winner if and only if $p$ is a winner
in $(C, S \cup T)$ when every voter in $T$ ranks the candidates from
right to left
by $c_{2k} > c_{2k-1} > \cdots > c_{\ell+2} > p > c_{\ell} > \cdots >
c_1$.

For the unique-winner case within scoring protocol $\alpha =
(\overbrace{1,\ldots,1}^{k},\overbrace{0,\ldots,0}^{k})$, if $p$'s
right neighbor in $L$ is tied with $p$ in $S$ (i.e.,
$\score_{S}(c_{\ell+2}) = \score_{S}(p)$), then set a voter of
lowest weight in $T$ to $p >
c_{\ell} > c_{\ell - 1} > \cdots > c_1 >
c_{\ell+2} > \cdots > c_{2k}$ and set all other voters in $T$ to rank
the candidates from right to left
by $c_{2k} > c_{2k-1} > \cdots > c_{\ell+2} > p > c_{\ell} > \cdots >
c_1$.  (Note that $T$ is single-peaked with respect to~$L$.)
Our algorithm can easily check whether this makes $p$ a unique winner
in $(C, S \cup T)$, and if so it accepts, otherwise it rejects.

\medskip
\noindent
{\bf Veto elections.}\quad
  The proof for veto elections (i.e.,
  for scoring protocols $\alpha = (1, \ldots, 1,0)$ where the number
  of candidates varies), similarly as the proof of the previous
  part
  of this theorem, hinges on
  the following observation: Given an instance of the constructive
  coalition weighted manipulation problem, where the votes are
  single-peaked with respect to the society's order~$L$, $p$ can be
  made a winner if and only if $p$ is the only candidate or
  $p$ is never last in~$S$.

For the unique-winner case, note that since at most two candidates
can be ranked last in single-peaked elections, unique winners
within veto elections can only exist if $\|C\| \leq 3$.  The result
for $\|C\| \in \{2, 3\}$ follows from the previous part of the theorem,
and the $\|C\| = 1$ case is trivial.
This completes the proof of the last part of Theorem~\protect\ref{t:fall-down}
and the whole proof.~\end{proofs}

We now come to an unusual case.  
For general votes (not limited to
single-peaked societies), the constructive coalition weighted
manipulation problem for
3-veto is in P for three or four candidates, and
is NP-complete
(and so resistant)
for five or more candidates
(see~\cite{hem-hem:j:dichotomy}).
(3-veto is not meaningfully defined for two or
fewer candidates.)
However, for single-peaked votes, 3-veto shows a remarkable behavior: Moving
from five to six candidates \emph{lowers} the complexity of this
manipulation problem.  In some papers,
authors
state that if one knows the number of candidates (if any) at which a system
switches from easy to manipulate to hard to manipulate,
it is easy by adding dummy candidates to see that
for all larger number of candidates the system remains hard.  The
following theorem should stand as a caution to take that view only if one
has carefully built and checked a ``dummy candidates'' construction for one's
specific case.

\begin{theorem}\label{t:toodles-csl}
For the single-peaked case, the constructive coalition weighted manipulation
problem (in both the unique-winner model and the
nonunique-winner model) for $m$-candidate 3-veto elections is in $\p$ for 
$m \in \{3,4,6,7,8,\ldots\}$ and is resistant (indeed, $\np$-complete) for
$m = 5$.\footnote{\singlespacing{}This
result holds both in our settled model where $L$, the society's ordering of
the candidates, is part of the input and in the model where $L$
is not given and the question is whether there is any societal ordering of
the candidates that allows the manipulators to succeed.}
\end{theorem}

\begin{proofs}
Consider the following four cases.
\begin{description}
\item[\boldmath{$m = 3.$}] In this case, we are looking at the scoring
  protocol $(0,0,0)$. It is immediate that in this scoring protocol
  all candidates are always tied for winner, and so $p$ can always be
  made a winner.  For the same reason, $p$ can never be a unique winner.

\item[\boldmath{$m = 4.$}] In this case, we are looking at the scoring
  protocol $(1,0,0,0)$. It is immediate that $p$ can be made a winner
  (a unique winner) if and only if $p$ is a winner (a unique winner)
  in the election where every manipulator ranks $p$ first.

\item[\boldmath{$m = 5.$}] In this case, we are looking at the scoring
  protocol $(1,1,0,0,0)$.  It is immediate that the constructive
  coalition weighted manipulation problem is in NP (simply guess
  single-peaked votes for the manipulators and verify that $p$ is a
  winner of the election).  To show NP-hardness, we reduce from 
  PARTITION.

  Given an instance of PARTITION, i.e., a set $\{k_1, \ldots, k_n\}$
  of $n$ distinct positive integers that sum to $2K$, construct the
  following instance of Constructive Coalition Weighted Manipulation:
  The set of candidates is $C = \{a,b,c,d,p\}$ with $p$ being our
  distinguished candidate.  The set of nonmanipulators $S$ consists of
  two voters, each of weight $K$.  One of the voters votes $c > a > p
  > b > d$ and the other voter votes $d> b > p > a > c$.  (Note that
  these two voters fix the society's order.)  We also have a set $T$ of
  $n$ manipulators.  The weights of the manipulators are $k_1, k_2,
  \ldots, k_n$.  We claim that there is a partition if and only if the
  manipulators can cast single-peaked votes that make $p$ a winner.

  First suppose that there exists a subset $S'$ of $\{k_1, \ldots,
  k_n\}$ that sums to $K$.  For every $i \in \{1, \ldots, n\}$, we set
  the weight $k_i$ manipulator to $p > a > b > c > d$ if $k_i$ in $S'$
  and to $p > b > a > c > d$ otherwise.  In the resulting election,
  $a$, $b$, and $p$ each score $2K$ points and $c$ and $d$ score $K$
  points. It follows that $p$ is a winner of the election.

  For the converse, suppose the voters in $T$ vote single-peaked such
  that $p$ is a winner of the election.  Since for every single-peaked
  vote, if $p$ gets a point then $a$ or $b$ gets a point, it follows
  that $\score_T(p) \leq \score_T(a) + \score_T(b)$.  Since $p$ is a
  winner of the resulting election, $\score_T(p) \geq K + \score_T(a)$
  and $\score_T(p) \geq K + \score_T(b)$.  It follows that
  $\score_T(p) = 2K$ and that $\score_T(a) = \score_T(b) = K$.  But
  then the weights of the voters in $T$ that give a point to $a$ sum
  to $K$ and so we have found a partition.

  Changing the weights of the voters in $S$ from $K$ to $K-1$ 
  handles the unique-winner case.

\item[\boldmath{$m \geq 6.$}] This has already been proven as the
  special case $k_1 \geq k_0 = 3$ of part~\ref{part:majority-ones} of
  Theorem~\ref{t:fall-down}.
\end{description}
This completes the proof of
Theorem~\protect\ref{t:toodles-csl}.~\end{proofs}

We now present some cases that are known
(see~\cite{hem-hem:j:dichotomy}) to be NP-hard in the general case and
that we can prove remain hard even in the single-peaked case.

\begin{theorem}\label{t:stays-hard}
For the single-peaked case, the constructive coalition weighted
manipulation problem (in both the unique-winner model and the
nonunique-winner model) is resistant (indeed, $\np$-complete) for the
following scoring
protocols:
\begin{enumerate}
\item\label{part:3-1-0}
$\alpha = (3,1,0)$.
\item\label{part:borda-four}
$\alpha = (3,2,1,0)$, i.e., 4-candidate Borda 
elections.\footnote{\singlespacing{}Both cases of Theorem~\ref{t:stays-hard}
hold both in our settled model where $L$, the society's ordering of
the candidates, is part of the input and in the model where $L$
is not given and the question is whether there is any societal ordering of
the candidates that allows the manipulators to succeed.}
\end{enumerate}
\end{theorem}

\begin{proofs}
  We consider the two scoring protocols, $\alpha = (3,1,0)$ and
  $\alpha = (3,2,1,0)$ separately. In each case it is easy to 
  see that the constructive coalition weighted manipulation problem
  is in $\np$ and it remains to give the $\np$-hardness proof.

\medskip
\noindent
{\bf The scoring protocol \boldmath{$\alpha = (3,1,0)$}.}\quad
  To show NP-hardness, we
  reduce from PARTITION.  Given a set $\{k_1, \ldots, k_n\}$
  of $n$ distinct positive integers that sum to $2K$, define the following
  instance of the constructive coalition weighted manipulation
  problem: The candidate set is $C = \{a, b, p\}$, there are two
  nonmanipulative voters in~$S$, one of the form $a > p > b$ and one
  of the form $b > p > a$ (note that $S$ is single-peaked  with
  respect to $a \,L\, p \,L\, b$, its reverse, and no other order) and both of weight $5K$, and there
  are $n$ manipulators in $T$ where the $i$th manipulator has
  weight~$k_i$, $1 \leq i \leq n$.

  We claim that there exists a partition (i.e., a set $A \subseteq I =
  \{1, 2, \ldots , n\}$ such that $\sum_{i \in A} k_i = \sum_{i \in
    I-A} k_i$) if and only if the manipulators can cast single-peaked
  votes (with respect to~$L$) that make $p$ a winner of $(C, S \cup
  T)$.

  From left to right, suppose there is a partition $A \subseteq I =
  \{1, 2, \ldots , n\}$ with $\sum_{i \in A} k_i = \sum_{i \in I-A}
  k_i$.  Set those voters in $T$ whose weights are~$k_i$, $i \in A$,
  to $p > a > b$, and set the remaining voters in~$T$ (i.e., those
  with weights~$k_i$, $i \in I-A$) to $p > b > a$.  Note that $T$ is
  single-peaked with respect to~$L$.  We have $\score_{S \cup T}(p) =
  10K + 6K = 16K$ and $\score_{S \cup T}(a) = \score_{S \cup T}(b) =
  15K + K = 16K$, so $p$ is a winner of $(C, S \cup T)$.

  From right to left, suppose $T$ is set such that it is single-peaked
  with respect to~$L$ and $p$ is a winner of $(C, S \cup T)$.  Without
  loss of generality, we may assume that $p$ is first in every voter
  in~$T$.  Thus $\score_{S \cup T}(p) = 16K$.
  So, for $p$ to be a winner, we must have
  $\score_{T}(a) \leq K$ and $\score_{T}(b) \leq K$.  But then $T$
  corresponds to a partition.
  
  Changing the weights of the voters in $S$ from $5K$ to $5K-1$ 
  handles the unique-winner case for $\alpha = (3,1,0)$.

\medskip
\noindent
{\bf The scoring protocol \boldmath{$\alpha = (3,2,1,0)$}.}\quad
  We again give a reduction from PARTITION to show $\np$-hardness.
  We start with the nonunique-winner case.

  Given an input to the PARTITION problem, i.e., a set
  $\{k_1, \ldots, k_n\}$ of $n$ distinct positive integers
  that sum to $2K$, we form the
  following instance of Constructive Coalition Weighted Manipulation.
  The candidate set is $C = \{a,b,c,p\}$ with $p$ being our
  distinguished candidate, the single-peakedness order
  $L$ is
  given by $a \, L \, b \, L \, p \, L \, c$,\footnote{\singlespacing{}If
    $L$ is not part of the input, we
    add two voters to $S$, namely $a > b > p > c$ and $c > p > b
    >a$, each of weight $1$, who fix the society's order to be
    our required $L$ (or, equivalently, $L$'s reverse).} and the set $S$ of
  nonmanipulative voters contains two voters, $v_1$ and
  $v_2$.  Voter $v_1$
  has weight $11K$ and preference order $c > p > b > a$ and voter
  $v_2$ has weight $7K$ and preference order $b > a > p > c$. There
  are $n$ manipulators in the manipulator set $T$ and the $i$th
  manipulator, $1 \leq i \leq n$, has weight $k_i$.  We claim that the
  manipulators can cast votes that ensure that $p$ is a winner in
  $(C,S\cup T)$ if and only if $\{k_1, \ldots, k_n\}$ can be
  partitioned into two subsets that sum to~$K$.

\begin{table}[tb!]
\begin{center}
\begin{tabular}{|l|l|p{12.0cm}|}
  \hline
  & \multicolumn{1}{c|}{Order} & \multicolumn{1}{c|}{Comments} \\
  \hline
  $o_1$ & $a > b > p > c$ & The manipulators are better off casting
  $o_5$ instead ($p$ gets more points, $a$ gets fewer points, and the
  scores of $b$ and $c$ do not change). \\
  \hline
  $o_2$ & $b > a > p > c$ & The manipulators are better off casting
  $o_1$ instead (and, in effect, are better off casting $o_5$) because
  $a$ cannot become a winner anyway and casting $o_1$ decreases $b$'s
  score. \\
  \hline
  $o_3$ & $b > p > a > c$ & The manipulators are better off casting
  $o_5$ instead ($p$ gets more points, $b$ gets fewer points, and the
  scores of $a$ and $c$ do not change). \\
  \hline
  $o_4$ & $b > p > c > a$ & The manipulators are better off casting
  $o_5$ instead ($p$ gets more points, $b$ gets fewer points, $c$ gets
  fewer points, $a$ gets more points but $a$ cannot become a winner
  anyway).\\
  \hline
  $o_5$ & $p > b > a > c$ & The manipulators might use this order. \\
  \hline
  $o_6$ & $p > b > c > a$ & The manipulators are better off casting
  $o_5$ instead ($c$ gets fewer points and $a$ gets more points but
  $a$ cannot become a winner anyway; the scores of $p$ and $b$ do not
  change).\\
  \hline
  $o_7$ & $p > c > b > a$ & The manipulators might use this order.\\
  \hline
  $o_8$ & $c > p > b > a$ & The manipulators are better off casting
  $o_7$ instead ($p$ gets more points, $c$ gets fewer points, and the
  scores of $a$ and $b$ do not change). \\
  \hline
\end{tabular}
\end{center}
\caption{\label{tab:borda}The eight preference orders consistent
  with $a \, L \, b \, L \, p \, L \, c$ and comments as to which of them
  the manipulators may choose to use.}
\end{table}

  Not counting the manipulators' votes, the candidates have the following
  scores:
  \begin{enumerate}
  \item $\score_S(a) = 14K$,
  \item $\score_S(b) = 32K$,
  \item $\score_S(p) = 29K$, and
  \item $\score_S(c) = 33K$.
  \end{enumerate}
  Since the total weight of the manipulators is $2K$ and we use
  scoring vector $(3,2,1,0)$, it is easy to see that irrespective of
  what votes the manipulators cast, candidate $a$ can never become a
  winner. Using this fact, it is easy to see that out of the eight
  preference orders $o_1, o_2, \ldots, o_8$ (see
  Table~\ref{tab:borda}) that are consistent with $L$ the manipulators
  can limit themselves to choosing between
  $o_5 = p > b > a > c$ and $o_7 = p > c > b > a$. (For each
  of the remaining orders Table~\ref{tab:borda} explains why either
  $o_5$ or $o_7$ is at least as good an order to cast.)

  Let $z_5$ be the total weight of the manipulators that vote $o_5$
  and let $z_7$ be the total weight of the manipulators that vote
  $o_7$. Clearly, $z_5 + z_7 = 2K$. Now, including the manipulators, 
  the candidates have the following scores:
  \begin{enumerate}
  \item $\score_{S \cup T}(a) = 14K + z_5$,
  \item $\score_{S \cup T}(b) = 32K + 2z_5 + z_7$,
  \item $\score_{S \cup T}(p) = 29K + 3z_5+3z_7 = 35K$, and
  \item $\score_{S \cup T}(c) = 33K + 2z_7$.
  \end{enumerate}
  It is easy to see that $p$ is a winner of election $(C,S\cup T)$ if and only
  if $32K + 2z_5 + z_7 \leq 35K$ and $33K + 2z_7 \leq 35K$, which is
  equivalent to
  \begin{eqnarray}
    2z_5 + z_7 &\leq& 3K, \label{equ:4-borda-1} \\
    2z_7       &\leq& 2K. \label{equ:4-borda-2} 
  \end{eqnarray}
  Since
  $z_5 + z_7 = 2K$, it is easy to see that the
  above two inequalities hold if and only if $z_5 = z_7 = K$. Thus,
  the manipulators can ensure that $p$ is a winner if and only if the
  set $\{k_1, \ldots, k_n\}$ can be partitioned into two subsets that
  both sum up to~$K$. Since the reduction is clearly computable in
  polynomial time, the proof for the nonunique-winner case is complete.

  To handle the unique-winner case, we change the weights of the
  voters in $S$ in the above construction as follows.  Voter~$v_1$,
  whose preference order is $c > p > b > a$, has now weight $11K - 3$,
  and voter~$v_2$, whose preference order is $b > a > p > c$, has now
  weight $7K - 2$.  Then the candidates have the following scores in
  $(C,S)$:
  \begin{enumerate}
  \item $\score_S(a) = 14K - 4$,
  \item $\score_S(b) = 32K - 9$,
  \item $\score_S(p) = 29K - 8$, and
  \item $\score_S(c) = 33K - 9$.
  \end{enumerate}
  Regardless of which votes the manipulators cast, $a$ still cannot
  win, which again leaves $o_5 = p > b > a > c$ (with total
  weight~$z_5$) and $o_7 = p > c > b > a$ (with total weight~$z_7$) as
  the only reasonable choices for the voters in~$T$.  Now, in election
  $(C, S \cup T)$, we have the following scores:
  \begin{enumerate}
  \item $\score_{S \cup T}(a) = 14K - 4 + z_5$,
  \item $\score_{S \cup T}(b) = 32K - 9 + 2z_5 + z_7$,
  \item $\score_{S \cup T}(p) = 29K - 8 + 3z_5+3z_7 = 35K - 8$, and
  \item $\score_{S \cup T}(c) = 33K - 9 + 2z_7$.
  \end{enumerate}
Thus, $p$ is a unique winner of election $(C,S\cup T)$ if and only if
$32K - 9 + 2z_5 + z_7 < 35K - 8$ and $33K - 9 + 2z_7 < 35K - 8$,
which again is equivalent to~(\ref{equ:4-borda-1})
and~(\ref{equ:4-borda-2}).  The remainder of the proof is as in the
nonunique-winner case.
This completes the proof of
Theorem~\protect\ref{t:stays-hard}.~\end{proofs}

In fact, we can extend the idea behind the proofs of 
part~\ref{part:borda-three} of Theorem~\ref{t:fall-down} and
part~\ref{part:3-1-0} of Theorem~\ref{t:stays-hard} to 
obtain the following dichotomy result for 3-candidate scoring
protocols in the single-peaked case.

\begin{theorem}\label{t:3-candidate-dichotomy}
Consider a 3-candidate scoring protocol, namely,
$\alpha = (\alpha_1, \alpha_2, \alpha_3)$,
$\alpha_1 \geq \alpha_2 \geq \alpha_3$,
$\alpha_1 \in \nat$, $\alpha_2 \in \nat$, 
$\alpha_3 \in \nat$.
For the single-peaked case, the constructive coalition weighted
manipulation problem (in both the unique-winner model and the
nonunique-winner model) is resistant (indeed, $\np$-complete) when
$\alpha_1  - \alpha_3 > 2 (\alpha_2 - \alpha_3) > 0$
and is in $\p$ otherwise.\footnote{\singlespacing{}This
result holds both in our settled model where $L$, the society's ordering of
the candidates, is part of the input and in the model where $L$
is not given and the question is whether there is any societal ordering of
the candidates that allows the manipulators to succeed.}
\end{theorem}

\begin{proofs}
Consider a 3-candidate scoring protocol $\alpha = (\alpha_1, \alpha_2,
\alpha_3)$, where $\alpha_1 \in \nat$, $\alpha_2 \in \nat$, $\alpha_3
\in \nat$, and $\alpha_1 \geq \alpha_2 \geq \alpha_3$.  
Without loss of generality, we may assume that $\alpha_3 = 0$.
If $\alpha_3$ were not $0$, we could replace our scoring protocol
with $(\alpha_1-\alpha_3, \alpha_2-\alpha_3, 0)$ (see~\cite{hem-hem:j:dichotomy}).
Suppose
that $\alpha_1 - \alpha_3 > 2 (\alpha_2 - \alpha_3) > 0$, 
i.e., $\alpha_1 > 2\alpha_2 > 0$.  We show
that in this case the constructive coalition weighted manipulation
problem is $\np$-complete.  Membership in $\np$ is immediate and
to prove $\np$-hardness we
reduce from PARTITION.  Given a PARTITION
instance, i.e., a set $\{k_1, \ldots, k_n\}$ of $n$ distinct positive
integers that sum to $2K$, we construct an instance of the
constructive coalition weighted manipulation problem as follows.  Let
$a$, $b$, and $p$ be the candidates in~$C$, where $p$ is the
distinguished candidate the manipulators want to make win.  There are
two nonmanipulators in~$S$, with preferences $a>p>b$ and $b>p>a$ and
both having weight $(2\alpha_1 - \alpha_2)K$. 
There are $n$ manipulators in~$T$, where the $i$th
manipulator has weight $(\alpha_1 - 2\alpha_2)k_i$, $1 \leq i \leq n$.  Note that, due to
$\alpha_1 > 2\alpha_2 > 0$, each voter has a
positive integer weight. 
We fix the society's order $L$ to be $a\, L\, p\, L \, b$. Note
that $L$ and its reverse are the only two orders consistent with the votes
in~$S$,
so even in the model where $L$ is not part of the input, we will still,
in effect, be working with $L$ as the society's order.

We claim that there exists a partition (i.e., a set $A \subseteq I =
\{1, 2, \ldots , n\}$ such that $\sum_{i \in A} k_i = \sum_{i \in I-A}
k_i$) if and only if the manipulators can cast single-peaked votes
(with respect to $L$) that make $p$ a winner of $(C, S
\cup T)$.

Suppose there is a partition, i.e., a set $A \subseteq I = \{1, 2,
\ldots , n\}$ such that $\sum_{i \in A} k_i = \sum_{i \in I-A} k_i$.
To make $p$ win, set $(\alpha_1 - 2\alpha_2)K$ 
vote weight of $T$ (say, each of the manipulators
corresponding to~$A$) to $p>a>b$ and the remaining manipulators to
$p>b>a$.  Note that $T$ is single-peaked with respect to $L$.
We have the following scores:
\begin{eqnarray*}
\score_{S \cup T}(p) & = &
  2 \alpha_2 (2 \alpha_1 - \alpha_2) K
+ 2 \alpha_1 (\alpha_1 - 2 \alpha_2) K
=  2 \left( \alpha_{1}^{2} - \alpha_{2}^{2} \right) K ; \\
\score_{S \cup T}(a) & = &
  \alpha_1 (2 \alpha_1 - \alpha_2) K
+ \alpha_2 (\alpha_1 - 2 \alpha_2) K
= 2 \left( \alpha_{1}^{2} - \alpha_{2}^{2} \right) K ; \\
\score_{S \cup T}(b) & = &
  \alpha_1 (2 \alpha_1 - \alpha_2) K
+ \alpha_2 (\alpha_1 - 2 \alpha_2) K
= 2 \left( \alpha_{1}^{2} - \alpha_{2}^{2} \right) K .
\end{eqnarray*}
So $p$ is a winner of election $(C, S \cup T)$.

Conversely, suppose the votes in $T$ can be cast such that $p$ is a
winner of $(C, S \cup T)$.  Note that
\[
\sum_{c \in C} \score_{S \cup T}(c) =
 6 \left( \alpha_{1}^{2}  - \alpha_{2}^{2} \right) K.
\]
So, in order for $p$ to be a winner, $p$'s score in $(C, S \cup T)$
needs to be at least $2 \left( \alpha_{1}^{2} -\alpha_{2}^{2}  \right) K$.  
This can happen only
if $p$ is ranked first by every voter in~$T$.  And if we set $T$'s votes like that,
$p$'s score is exactly $2 \left( \alpha_{1}^{2} -\alpha_{2}^{2} \right) K$.  
In order for $p$ to be
a winner, $a$'s score and $b$'s score need to be $2 \left(
\alpha_{1}^{2} - \alpha_{2}^{2} \right) K$ each.  This means that the vote weight of the voters in
$T$ voting $p>a>b$ is equal to the vote weight of the voters in $T$
voting $p>b>a$.  Thus there exists a partition.

Since the reduction is polynomial-time computable, the constructive
coalition weighted manipulation problem for the scoring protocol
$\alpha = (\alpha_1, \alpha_2, 0)$ is $\np$-complete, provided
that $\alpha_1 > 2\alpha_2  > 0$.

The unique-winner case can be handled by multiplying all weights in
the above construction by a suitable constant (depending on 
$\alpha_1$ and $\alpha_2$) and then subtracting $1$ 
from the weights in~$S$. 
The idea is that subtracting 1 from the weights in~$S$ ensures that
$p$ can become the unique winner if there is a partition, and multiplying
the weights, prior to the subtraction, ensures that
the subtraction does not have any 
side effects that would
invalidate the (analogue of the) above correctness argument
in this modified reduction.

Now suppose $\alpha_1 \geq \alpha_2 \geq 0$ but it is
\emph{not} the case that $\alpha_1 > 2 \alpha_2 > 0$.  We show
that in this case the constructive coalition weighted manipulation
problem is in~$\p$.
If $\alpha_2 = 0$ then our scoring protocol is equivalent to
plurality for three candidates, and we accept if and only if
$p$ is a winner (a unique winner) when all the manipulators
rank $p$ first. Let us now focus on the case where $\alpha_2 > 0$,
and so, by assumption, $\alpha_1 \leq 2\alpha_2$.
Without loss of generality, it is enough to consider the following two
orderings $L$ witnessing the society's single-peakedness.  (Again, if
$L$ is not part of the input, we simply try all societal orderings.)
We'll handle the nonunique-winner case and the unique-winner case in
parallel.

\begin{description}
\item[Case~1:] $a \,L\, b \,L\, p$.
  In this case, if the goal is to make $p$ a winner (a unique winner),
  a best vote for all manipulators in $T$ is clearly $p > b > a$.

\item[Case~2:] $a \,L\, p \,L\, b$.  To be consistent with $L$, the
  votes allowed in $S$ are $p > a > b$, $p > b > a$, $a > p > b$, and
  $b > p > a$ (see
  the proof of
  part~\ref{part:borda-three} of Theorem~\ref{t:fall-down}).
  Due to which votes are allowed in~$S$, we have $2 \cdot \score_S(p)
  \geq \score_S(a) + \score_S(b)$.  It follows that $\score_S(p) \geq
  \score_S(a)$ or $\score_S(p) \geq \score_S(b)$.  If we are in the
  nonunique-winner model, set all voters in $T$ to $p>a>b$ if
  $\score_S(p) \geq \score_S(a)$, and to $p>b>a$ if $\score_S(p) \geq
  \score_S(b)$.  If we are in the unique-winner model
  and $\alpha_1 > \alpha_2$, set all voters in $T$ to $p>b>a$ if
  $\score_S(a) \geq \score_S(b)$ and to $p>a>b$ otherwise.  The
  $\alpha_1 = \alpha_2$ case follows from part~\ref{part:majority-ones}
  of Theorem~\ref{t:fall-down}.
\end{description}
In both cases, we accept if the manipulation action described above
was successful (i.e., made $p$ a winner in the nonunique-winner case
(a unique-winner in the unique-winner case) in election $(C, S \cup
T)$), and we reject otherwise.~\end{proofs}

Finally, we mention that despite the NP-hardness results of
Theorems~\ref{t:stays-hard} and~\ref{t:3-candidate-dichotomy},
in all these cases there are polynomial-time manipulation algorithms for
the case when the candidate the coalition wants to win is either the top or
the bottom candidate in society's input linear order on the candidates.

\section{Related Work}
\label{sec:related-work}

The paper that inspired our work is Walsh's ``Uncertainty in
Preference Elicitation and
Aggregation''~\cite{wal:c:uncertainty-in-preference-elicitation-aggregation}.
Among other things, in that paper he raises the issue of manipulation
in single-peaked societies.  Our paper follows his model of assuming
society's linear ordering of the candidates is given and that
manipulative voters must be single-peaked with respect to that
ordering.  However, our theme and his differ.  His manipulation
results present cases where single-peakedness leaves an
$\np$-completeness shield intact.  In particular, for both the
constructive and the destructive cases, he shows that the coalition
weighted manipulation problem for the single transferable vote
election rule for three or more candidates remains $\np$-hard in the
single-peaked case.  Although our Theorem~\ref{t:stays-hard} follows
this path of seeing where shields remain intact for single-peaked
preferences, the central focus of our paper is that single-peaked
preferences often remove complexity shields on manipulation and
control.  Walsh's paper for a different issue---looking at incomplete
profiles and asking whether some/all the completions make a candidate
a winner---proves both $\p$ results and $\np$-completeness results.
We're greatly indebted to his paper for raising and exploring the
issue of manipulation for single-peaked electorates.

As mentioned in the main text, Bartholdi and
Trick~\cite{bar-tri:j:stable-matching-from-psychological-model},
Doignon and Falmagne~\cite{doi-fal:j:unidimensional}, and 
Escoffier, Lang, and
{\"O}zt{\"u}rk~\cite{esc-lan-ozt:c:single-peaked-consistency} have
provided efficient algorithms for testing single-peakedness and
producing a valid candidate linear ordering, for the case when votes
are linear orders. 

Other work is more distant from our work but worth mentioning.
Conitzer~\cite{con:j:eliciting-singlepeaked} has done an
interesting, detailed study showing that (in the model where votes are
linear orders) preferences in single-peaked societies can be quickly
elicited via comparison queries (``Do you prefer candidate~$i$ to
candidate~$j$?'').  He studies the case when the linear order of
society is known and the case when it is not.  We mention in passing
that we have looked at the issue of preference elicitation in
single-peaked societies (where the linear order is given) of approval
vectors via approval queries (``Do you approve of candidate~$i$?'').
It is immediately obvious that single-peakedness gives no improvement
for approval vectors and 1-approval vectors (approval vectors with
exactly one 1; this is a vote type and should not be confused with
``1-approval'' as it would be used in scoring systems, where the actual vote
is a linear order).  But for $j$-candidate $k$-approval vectors
(approval vectors with exactly $k$ 1's), $k \geq 1$, it is easy to see
that the general-case elicitation query complexity is exactly $0$ when
$j=k$ and is exactly $j-1$ when $j>k$, but for the single-peaked case,
the elicitation query complexity for $ik$-candidate $k$-approval
vectors is at most $i-1 + \left\lceil \log_2 k \right\rceil$.  That
is, we get a savings of a multiplicative factor of about $k$ from
single-peakedness.

Single-peaked preferences of course have been studied extensively in
political science.  We in particular mention that Ballester and
Haeringer~\cite{bal-har:m:characterization-single-peaked} provide a
precise mathematical characterization of single-peakedness, that
Lepelley~\cite{lep:j:single-peaked} shows that single-peakedness
removes some negative results about the relationship between scoring
protocols and Condorcet-type criteria, and Gailmard, Patty, and
Penn~\cite{gai-pat-pen:btoappear:arrow-on-single-peaked-domains}
discuss Arrow's Theorem on single-peaked domains.  We refer the
interested reader also to the coverage of single-peaked preferences in
the excellent textbook by Austen-Smith and
Banks~\cite{aus-ban:b:positive-political-II}.

\section{Conclusions and Future Directions}
\label{sec:conclusions}

The central point of this paper is that single-peaked preferences
remove many complexity-theoretic shields against control and
manipulation.  That is, we showed that those shields, already under
frequency-of-hardness and approximation attacks from other quarters,
for single-peaked preferences didn't even exist in the first place.
It follows that when choosing election systems for electorates one
suspects will be single-peaked, one must not rely on results for those
systems that were proven in the standard, unrestricted preference
model.

This paper's work suggests many directions for future efforts.  For
single-peaked manipulation of scoring protocols, we gave some
$\np$-complete cases and some $\p$ cases.  But for manipulation of
scoring protocols (in the general model where society is not required
to be single-peaked), Hemaspaandra and Hemaspaandra
(\cite{hem-hem:j:dichotomy}, see
also~\cite{pro-ros:j:juntas,con-lan-san:j:few-candidates})
have provided a dichotomy theorem clearly classifying each case as
$\np$-complete or in~$\p$.  Can we obtain a dichotomy theorem for
manipulation of scoring protocols in single-peaked societies?
Theorem~\ref{t:3-candidate-dichotomy} achieves this for 
the case of three candidates.

Throughout this paper, single-peaked has meant the unidimensional
case.  Do the shield removals of this paper hold in, for example, an
appropriate two-dimensional (or $k$-dimensional) analogue?  (We mention
in passing that every profile of $n$ voters voting by linear orders
can be embedded into $\mathbb{R}^n$ in such a way that each voter and
candidate is a point in $\mathbb{R}^n$ and each voter prefers $c_i$ to
$c_j$ if her Euclidean distance to $c_i$ is less than to~$c_j$.)

In a human society with a large number of voters, even if one
issue, e.g., the economy, is almost completely polarizing the society,
there are bound
to be a few voters whose preferences are shaped by quite different
issues, e.g., a given candidate's religion.  So it would be natural
to ask whether the shield-evaporation results of this paper can be
extended even to societies that are ``very nearly'' single-peaked
(see~\cite[Section~6]{esc-lan-ozt:c:single-peaked-consistency}
and~\cite[Section~6]{con:j:eliciting-singlepeaked}
for discussion of nearness to single-peakedness in other contexts).

Finally, we mention that 
subsequent work of Brandt et
al.~\cite{bra-bri-hem-hem:ctoappear:single-peaked-bribery} further
explores single-peaked electorates---looking at bribery and at control by
partition of voters, and 
generalizing
Theorem~\ref{t:3-candidate-dichotomy} to each fixed number of candidates.

\paragraph*{Acknowledgments} 
For many helpful comments, discussions, and suggestions,
we are grateful to Steven Brams, 
Felix Brandt, Markus Brill, Felix Fischer, Paul
Harrenstein, J\'{e}r\^{o}me Lang, and Ariel Procaccia, 
to \emph{Information and Computation} editor Christos 
Papadimitriou, and to the anonymous TARK-09
referees.

\bibliography{gry-sp1-additions,gry-pf-sp1}
\bibliographystyle{alpha}

\end{document}